\documentclass[aps,twocolumn]{revtex4-1}
\usepackage{pdfpages}
\makeatletter
\AtBeginDocument{\let\LS@rot\@undefined}
\makeatother
\usepackage{graphicx}
\usepackage{xcolor}
\usepackage[english]{babel}
\usepackage{amsmath}
\usepackage{amssymb}
\usepackage{bm}
\usepackage{multirow}
\usepackage{makecell}
\usepackage{mathrsfs}
\usepackage{array}
\usepackage[normalem]{ulem}
\usepackage{pgffor}

\definecolor{darkgreen}{RGB}{0, 150, 0}

\def\RR{\mathscr{R}}

\def\N{{\mathbb N}}

\def\Ee{{\cal E}}
\def\Fe{{\cal F}}
\def\E{E}

\def\1{{\mathds{1}}}

\def\bR{{\bm r}}
\def\bZ{{\bm Z}}

\begin{document}

\author{Alice E. A. Allen}
\email{aa840@cam.ac.uk}
\affiliation%
{Engineering Laboratory, University of Cambridge, Trumpington Street,Cambridge, CB2 1PZ, United Kingdom}

\author{Genevi\`eve Dusson}
\email{genevieve.dusson@math.cnrs.fr}
\affiliation%
{Universit\'e Bourgogne Franche-Comt\'e, Laboratoire de Math\'ematiques de Besan\c{c}on, UMR CNRS 6623, Besan\c{c}on, France}

\author{Christoph Ortner}
\email{c.ortner@warwick.ac.uk}
\affiliation%
{Mathematics Department, University of British Columbia, 1984 Mathematics Rd, Vancouver, BC V6T 1Z2, Canada}

\author{G\'abor Cs\'anyi}
\email{gc121@cam.ac.uk}
\affiliation%
{Engineering Laboratory, University of Cambridge, Trumpington Street,Cambridge, CB2 1PZ, United Kingdom}

\title[]{Atomic Permutationally Invariant Polynomials for Fitting Molecular Force Fields}

\begin{abstract}
We introduce and explore an approach for constructing force fields for small molecules, which combines intuitive low body order empirical force field terms with the concepts
of data driven statistical fits of recent machine learned potentials. 
We bring these two key ideas together to bridge the gap between established empirical force fields that have a high degree of transferability on the one hand, and the machine learned potentials that are systematically improvable and can converge to very high accuracy, on the other. 
Our framework extends the atomic Permutationally Invariant Polynomials (aPIP) developed for elemental materials in [\textit{Mach. Learn.: Sci. Technol.} 2019 \textbf{1} 015004] to molecular systems. 
The body order decomposition allows us to keep the dimensionality of each term low, while the use of an iterative fitting scheme as well as regularisation procedures improve the extrapolation outside the training set. 
We investigate aPIP force fields with up to generalised 4-body terms, and examine the
performance on a set of small organic molecules.
We achieve a high level of accuracy when fitting individual molecules, comparable to those of the many-body machine learned force fields. 
Fitted to a combined training set of short linear alkanes, the accuracy of the aPIP force field still significantly exceeds what can be expected from classical empirical force fields, while retaining reasonable transferability to both configurations far from the training set and to new molecules.
\end{abstract}

\maketitle

\section{Introduction}
Molecular mechanics (MM) with classical empirical force fields has been used to perform simulations of organic molecules for many decades~\cite{Jorgensen1988, Weiner1984}.  
One of the principle reasons why such force fields have been so successful is that the simplicity of their functional form results in both a low body order and relatively few fitting parameters. 
This allows the parameters to be fit using just a small amount of quantum mechanical (QM) or experimental data, and the problems associated with overfitting do not readily occur. The simple, chemically intuitive functional form makes the force fields highly transferable, giving reasonable results for molecules and conformations far away from those that were used to fit the parameters. 
Over time, improvements were made in the description of both the intermolecular interactions, particularly through the construction of polarizable models~\cite{Ponder2010}, and the intramolecular interactions, mainly with the development of Class II force fields~\cite{Maple1994,Ewig2001, Sun2016} that introduced new couplings between bond and angle terms. However, despite these developments the accuracy of classical force fields remains limited by the restrictive functional forms, the very same that gave rise to their success. 
This is particularly noticable when having to make unavoidable trade-offs between the accuracy of different observables. Freeing up the functional form has been achieved for only low body orders (up to three-body) e.g. in the ChIMES force field~\cite{chimes1,chimes2}. 

Over the past few years, a completely new direction emerged: the development of machine learning (ML) based potentials~\cite{Behler2007,bartok2010gaussian} has led to a significant improvement in accuracy for small molecules~\cite{Behler2011, Rupp2012, Bartok2013, Bartok:2013gf, Behler2015, Manzhos2015, Behler2016, Bartok:2017hz, Smith2017,Chmiela:2017ff, Chmiela2018, Veit:2019gp}. 
It has been demonstrated that these new ML models can perform molecular dynamics (MD) simulations in some cases and be transferred to some extent to molecules they have not been explicit fit to~\cite{Smith2019}. 
The long term goal of these developments is to obtain a general model  having an accuracy comparable to accurate ab initio methods such as CCSD(T)~\cite{Raghavachari1989-cx}, and a speed and transferability comparable to classical force fields.
The various formalisms of the recent ML models represent, on the surface, a radical departure from that of empirical force fields. The aim of this paper is to bridge this formalism gap, and to seek answers to questions such as: what makes the ML models accurate, is it their high dimensionality (i.e. body order) or their flexible functional form? How much additional accuracy is gained by allowing a controlled increase in body order?

\begin{figure}[!ht]
    \centering
    \includegraphics[width=3.2in]{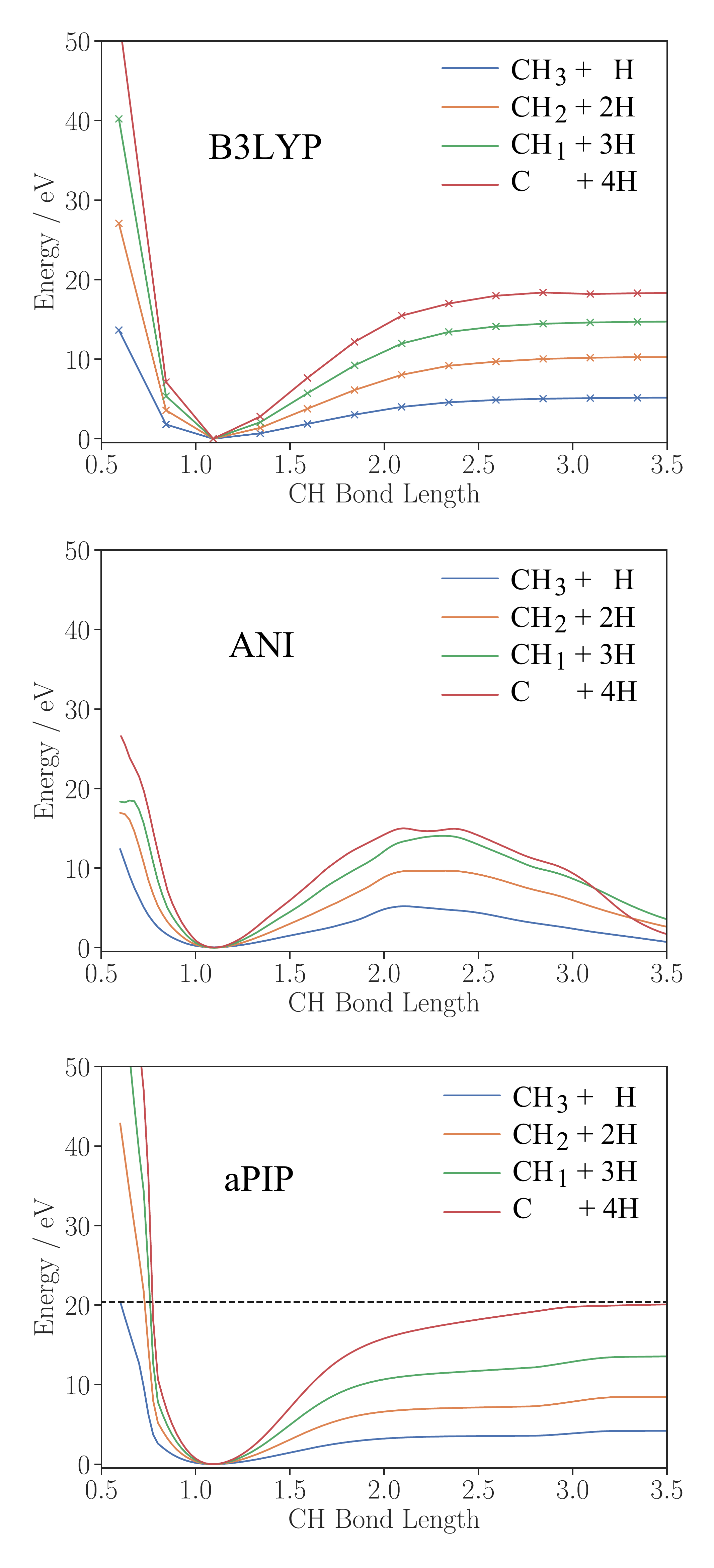}
    \caption{\label{fig:ch4dis}Removing some number of hydrogens from methane using DFT/B3LYP (top), ANI (middle) and aPIPs (bottom). Both ANI and the aPIP  were fitted only to near equilibrium geometries (the aPIP shown here was fitted to a database of small linear alkanes, including distortions of bond lengths up to 0.3~\AA). 
    }
\end{figure}

Predating the recent surge of interest in machine learned force fields, there is a significant literature for accurate fitting of molecular potential energy surfaces (PES)~\cite{Behler2016, Deringer2019-ss}. 
One of the most prominent and {\em systematic} approaches is due to Bowman and Braams~\cite{Braams2009-wi,Huang:2005, Xie:2005,Qu2019-vb,Nandi2019-bl}. 
Given a fixed molecular composition, a basis consisting of permutationally invariant polynomials (PIPs) of the interatomic distances is constructed and then used to approximate the PES by least squares fitting. 
Models using PIPs have very high accuracy, below 1~meV/molecule, and successfully reproduced the properties of a number of small molecules including $\mathrm{CH_5^+}$, $\mathrm{H_2O}_2$ and malonaldehyde~\cite{Braams2009-wi,Huang:2005, Xie:2005}. 
They have also been used as building blocks in models of water such as MBpol by Paesani et al~\cite{Babin:2012kv,Babin:2013fs,Babin:2014bn,Medders:2014bp}, probably the most accurate water model to date.

The difficulty in extending the PIP formalism to larger molecules or larger clusters of small molecules is in the scaling of the computational cost with the number of atoms. The number of permutationally invariant polynomials that are used as the basis grows factorially, in fact we were unable to obtain basis functions for six atoms of the same kind with full permutation symmetry. One way out of this crushing scaling is to employ a so-called ``fragmentation scheme''~\cite{Qu2019-vb}, in which parts of the molecule are explicitly grouped into fragments, and permutations between the groups are excluded from the symmetries of the basis. However, the choice of the fragments, based on distances in the initial molecule template, is manual and therefore limited to molecules which can be clearly split into suitable fragments. Potentials for large cyclic molecules or flexible molecules cannot be produced with this method.

\begin{figure*}[!ht]
    \centering
    \includegraphics[width=7in]{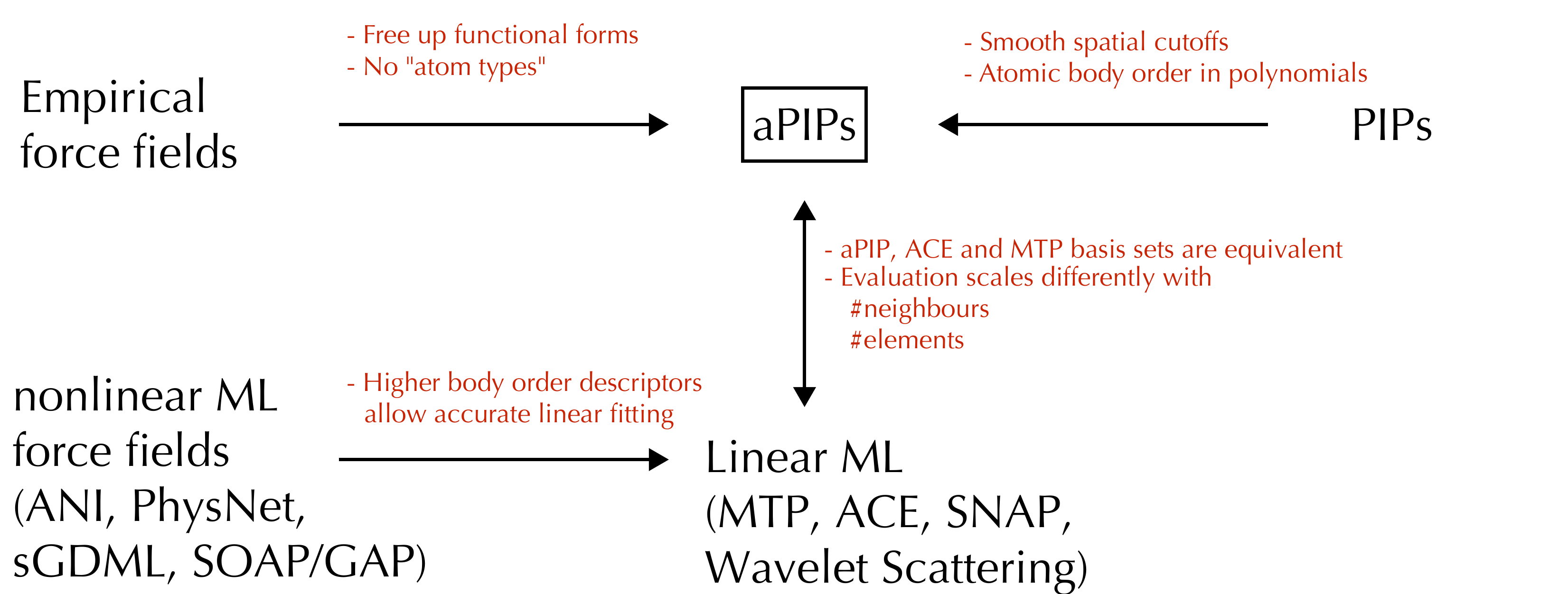}
    \includegraphics[width=7in]{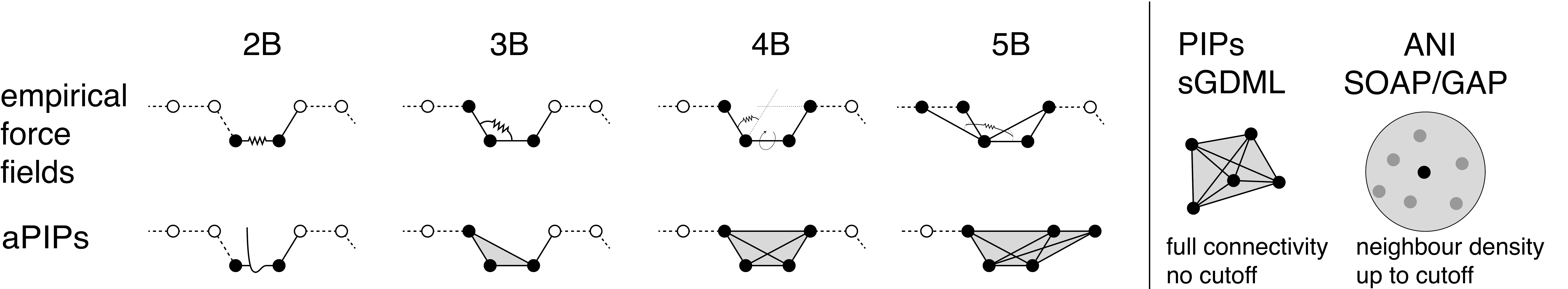}
    \caption{\label{fig:forcefields}Top: Relationships between different force field schemes. Bottom: a graphical representation of the body order terms in empirical force fields and in the aPIP framework. The shading of the 3-,4- and 5-simplices emphasises that full dimensionality of the simplex is represented, and thus the $n=$ 3-, 4-, and 5-body terms are 3, 6 and 9 dimensional functions, respectively (given by $3n-6$). 
    On the right, the cartoons represent (i) the PIP method of Bowman \& Braams~\cite{Braams2009-wi} and sGDML~\cite{Chmiela2019} that both describe a whole molecule always using its full dimensional representation, and (ii) atom-centered ML methods~\cite{Smith2017,Bartok2015-xc} that  represent the complete environment of an atom up to a cutoff. All of these approaches are many-body and do not take advantage of the atomic body-ordered decomposition in the way empirical force fields and aPIPs do. 
    }
\end{figure*}

A little over a decade ago, a new approach to making potentials was devised, inspired by the developments in computer science. 
Instead of systematically expanding the potential energy function using an analytically defined basis set, or using carefully designed chemically intuitive descriptions of atomic interactions, the idea was to describe the {\em entire local neighbourhood} of an atom using a set of descriptors, and then use nonlinear regression techniques to fit a model with thousands of parameters to first principles data. 
First, Behler and Parrinello introduced atom-centered symmetry functions and used a feed-forward artificial neural networks~\cite{Behler2007}; later, spherical harmonics were used in combination with Gaussian process (kernel) regression~\cite{bartok2010gaussian}. These new approaches led to exquisitely accurate interatomic potentials for strongly bound materials~\cite{Bartok:2013gf,Behler:2011it,Bartok:2017hz,Behler:2017,Bartok:2018ih} and condensed phase molecular materials~\cite{Bartok2013, Morawietz2013,Gastegger2016, Nguyen2018-le, Veit:2019gp}.
Although using neural networks to fit potentials was not itself new (e.g. see Refs.~\citenum{Manzhos2006,Ho2016}) these aforementioned models had finite interaction range, resulting in linear scaling cost, which opened the door to simulations of large systems with unprecedented accuracy. 

The field has since blossomed, with many novel approaches introduced~\cite{ Manzhos2015, Thompson2015, Shapeev:2016kn, Kolb2016, Huang2017, Schutt2017, Yao2018, Unke:2019bp}. 
Particularly notable is the ANI series of models for organic molecules~\cite{Smith2017, Smith2018-less, Smith2019}. 
For moderate sized molecules and small clusters very accurate custom made (non-transferable) models can be made using simply the interatomic distances as the set of descriptors and a Gaussian kernel~\cite{Bartok2013, Bartok:2013gf, Chmiela:2017ff, Chmiela2018, Cole2019}. 
Along with these successes, however, due to the high dimensional nature of these fits, come the problems of extrapolation and transferability. 
All these ML models (both for materials and molecules) are guaranteed to be accurate only for configurations quite near their training set, and can become uncontrolled or even nonsensical far from there. 
The practice of making ML models has therefore largely focused on how to create suitable (and rather large) training sets, how to detect when the model goes ``out of scope'' etc. 
Such problems do not exist for the empirical force fields: although their accuracy is only moderate, and not systematically improvable, they never give catastrophically incorrect results. They behave much more reasonably in extrapolation, even without explicit fitting to such configurations, since chemical intuition is built in through the functional form and leads to much lower-dimensional objects to fit. 

An illustration of such a problem is in Fig.~\ref{fig:ch4dis}, showing the ANI model\cite{Smith2017} behaving unphysically as hydrogens are removed from a methane molecule. The reason is simple: neither such bond dissociations, nor isolated atoms were included in its fitting database; no doubt if they were, the results would look much better. We emphasise that this is no indictment of the ANI model in particular, and we expect all direct high dimensional fits to behave similarly, or worse.

In this work, we introduce a way of making force fields for molecules that has the transferability and reasonable extrapolation property, due to limited body order, of empirical force fields, and also the accuracy of the recent ML models, due to its systematic nature.  
The construction, which we call {\em atomic permutationally invariant polynomials} (aPIP), builds heavily on the PIP framework of Bowman and Braams, and generalises our earlier work for strongly bound materials Ref.~\citenum{Van_der_Oord2019-td}. We show that the ``best of both worlds'' is possible: chemically sensible functions leading to smooth dissociation curves, as shown in Fig~\ref{fig:ch4dis}, without compromising on the convergence and ultimate accuracy of the fit. Before we introduce the construction of multi-element aPIPs in detail, it is helpful to note that they can be viewed in two ways.

When developing empirical force fields, adding terms to the functional form is a natural way to improve a potential's accuracy, as was shown in the development of Class II force fields~\cite{Maple1994, Sun2016,Ewig2001}. 
These have additional cross-terms between the bond, angle and dihedral components and include higher degree terms~\cite{Maple1994}. 
This indeed improves the accuracy of the force field~\cite{Tsai2003} but these additional higher body order terms  introduce only very few new degrees of freedom. 
Manually adding such terms to the functional form becomes increasingly tedious and complex.  The aPIP construction can be viewed as a general framework to implement this idea of systematically increasing to arbitrary body order, while allowing for the full dimensional freedom at each order.  
By increasing the degree of the polynomials, the functional form becomes increasingly complex as higher degree bond and angle terms, as well as all possible cross-terms between angles and bonds, are automatically included.  
By retaining the atom-wise body ordered description, the dimensionality of the representation is kept small, as is the case for empirical force fields. (In fact existing empirical force fields can all be represented as aPIPs.) 
This is the crucial way in which aPIPs differ from the regular PIPs of Bowman and Braams, and brings us to the second, complementary view of aPIPs. 
They can instead be viewed as a version of PIPs in which we limit the combinatorial explosion in the number of terms with system size by (i) explicitly limiting the body order of the potential, and (ii) using a smooth spatial cutoff.
The latter can be seen as an automated way to implement the aforementioned ``fragmentation'' process, and also brings an atom centered view. 
In fact, it can be shown that in the limit of high body order, the aPIP basis is equivalent to the high dimensional ML approaches, and particularly closely linked to MTP~\cite{Shapeev:2016kn} and ACE~\cite{Drautz:2019, Bachmayr2019-ec}.
Figure~\ref{fig:forcefields} shows the conceptual relationships between many of the approaches mentioned in this section. The key advance with aPIPs is the ability to gradually and systematically increase the body order, describing each term in its full generality.  

Two more short points are in order. Firstly, due to the smooth cutoff we introduce, discrete atom types, as used in empirical force fields, are no longer necessary, although could still be used if desired. Secondly, just as empirical force fields have separate short and long range interaction terms, so too can aPIPs. We focus in this paper on the short range intramolecular interactions. Any existing long range model, whether describing electrostatics of van der Waals dispersion, can be added if desired.  

The construction of the aPIP potential  is as follows. To start with, the total energy is decomposed as a sum of body-ordered terms, each of which depends on the chemical element of the elements involved. More precisely, let us consider a system containing a total of $M$ atoms of $K$ different elements $(Z_1,\ldots,Z_K)$ with positions and elements $\RR = \left((\bR_1,z_1),\ldots,(\bR_M,z_M)\right)$.
The total energy is decomposed as
\begin{widetext}
\begin{align}
   & \E(\RR) = \sum_{1\le k \le K} \sum_{\substack{i, \; \text{s.t.} \\  z_{i} = Z_k}} 
   \E_1^{(Z_k)}(\bR_{i})  %
      +  \sum_{1\le k_1 \le k_2 \le K} \sum_{\substack{i_1 \neq i_2 \; \text{s.t.} \\ (z_{i_1},z_{i_2}) = (Z_{k_1},Z_{k_2})}}
      \hspace{-.6cm}
      \E_2^{(Z_{k_1},Z_{k_2})}(\bR_{i_1}, \bR_{i_2}) \quad+ \notag\\
      &  +  \sum_{1 \le k_1 \le k_2 \le k_3 \le K}
      \sum_{\substack{i_1 \neq i_2 \neq i_3 \; \text{s.t.} \\ (z_{i_1},z_{i_2},z_{i_3}) \\ = (Z_{k_1},Z_{k_2},Z_{k_3})}}
      \hspace{-.8cm}
      \E_3^{(Z_{k_1},Z_{k_2},Z_{k_3})}(\bR_{i_1}, \bR_{i_2}, \bR_{i_3}) %
       + \cdots + %
      \sum_{1\le k_1 \le \ldots \le k_N \le K}
      \hspace{-.015cm}
      \sum_{\substack{i_1 \neq \dots \neq i_N \; \text{s.t.} \\
      (z_{i_1},\ldots,z_{i_N}) \\ = (Z_{k_1},\ldots,Z_{k_N})}}
      \hspace{-.6cm}
      \E_N^{(Z_{k_1},\ldots,Z_{k_N})}\big(\bR_{i_1}, \dots, \bR_{i_N}).\label{eq:bo-expansion}
\end{align}
\end{widetext}

Thus, the total energy is viewed as a sum of one-body, two-body, three-body contributions and so on, each body-order being itself described by many independent functions, separated with respect to the chemical elements.
In this paper, we consider this expansion up to body-order 4, which keeps the dimensionality of the potential low (up to 6 dimensions for four-body terms).
To guarantee the rotation and permutation invariance of the global PES, we enforce the symmetries at each body-order and for each component $E_n^{(Z_{k_1},\ldots,Z_{k_n})}$ that we denote by $E_n^{\bZ}$, combining the element indices into a vector. As detailed in the next section, we transform the cartesian coordinates in each term into interatomic distances and angle variables, which are rotation-invariant. We then construct permutation-invariant polynomials of these variables following Ref.~\citenum{Braams2009-wi}. Finally, these polynomials are globally fit using a linear least-squares fit to energy and force data. In order to limit the evaluation cost and be able to treat large molecules, we employ a distance-based cutoff, restricting the sums in each term of the body-order expansion \eqref{eq:bo-expansion} to nearby atoms. Furthermore, in order to avoid the presence of holes in the PES, i.e. very large negative values of the energies for some reasonable physical configurations, we add {\em regularisation} to the least-squares fit, thus improving the smoothness of the potential. An iterative data gathering and fitting procedure is also used to eliminate holes in the PES. Such techniques are essential for the potential to be readily used for a wide range of systems and applications. 

In this work, we set out to explore the use of aPIPs, rather than introduce a specific force field parametrisation for future use, and thus focus only on a handful of small organic molecules made of a few different elements. 
Note that the approach is well suited to applications with many chemical elements, due to its inherently favourable scaling: each cluster appears once and only once in the expansion of the energy, so that the overall evaluation cost of the potential is independent of the number of distinct elements.

\section{Theory}

\subsection{Symmetric polynomial basis}

We introduce a basis of permutation and rotation invariant polynomials for fitting molecular PES, extending the construction of~\mbox{Ref.~\citenum{Van_der_Oord2019-td}} to treat multi-element systems. While our focus is on molecules, our construction directly applies to multi-component alloys.
As in \mbox{Ref.~\citenum{Van_der_Oord2019-td}}, the starting assumption is that the body-order expansion~\eqref{eq:bo-expansion} can be truncated at a moderate to low body-order $N$ to obtain an accurate PES. 
The body-ordered components $E_n^\bZ$ are then constructed to incorporate rotation symmetry and permutation invariance with respect to identical particles.
The main difference from Ref.~\citenum{Van_der_Oord2019-td} concerns the invariant polynomials used, which are adjusted to the permutation groups considered. Differences of our method from the original PIPs~\cite{Braams2009-wi} are discussed in detail in the Introduction, see also below. 

\subsubsection{Rotation-invariant coordinates}

Given a body-order $n$ and atomic positions $(\bR_1,\ldots,\bR_n)$, we can define rotation-invariant (RI) coordinates in two different ways:
\begin{itemize}
    \item[(D)] {\em Distance-based coordinates:} Let $u_{ij}$ denote a distance transform, e.g., $u_{ij} = r_{ij}$, $u_{ij} = e^{- \alpha r_{ij}},$ or inverse distance variables $u_{ij} = r_{ij}^{-p}$. Then we rewrite $E_n^{\bZ}$ as
    \[
        E_n^\bZ(\{\bR_i \}_{i=1}^n )
        = 
        E_n^{\bZ,\rm D}(\{u_{ij} \}_{1\leq i<j\leq n} ).
    \]
    The potentials proposed by Bowman \& Braams~\cite{Braams2009-wi} also employ 
    distance-based coordinates.
    \item[(DA)] {\em Distance-angle coordinates:} Particularly for molecules it is natural to consider {\em bond-angles}, which suggests using distance and angle variables. Given a center atom $i$, we define
    \[
        w_{jik} = \cos(\theta_{jik}) = \hat\bR_{ij} \cdot \hat\bR_{ik}.
    \]
     The combined distance and angle variables are  
    \[
        \{u_{ij}\}_{\substack{1\le j\le n \\ j\neq i}}, \{w_{jik}\}_{\substack{1\le j<k \le n \\  j,k\neq i}}.
    \]
    and the term $E_n^{\bZ}$ is rewritten as 
\[
        E_n^{\bZ}(\{\bR_i \}_{i=1}^n )
        = 
        E_n^{\bZ,\rm DA}(\{u_{1j}\}_{j=2}^n, \{w_{1jk}\}_{2\le j<k \le n} ).
\]
\end{itemize}
While distance-based coordinates were used e.g. in Refs.~\citenum{Braams2009-wi, Van_der_Oord2019-td}, and is the norm when working with PIPs~\cite{Babin:2012kv, Medders:2014bp}, we will focus on distance-angle coordinates, which happen to lead to better numerical results in the present context, as is illustrated in the Supplementary Information~S1.

\subsubsection{Permutation-invariant polynomials}
Given rotationally invariant coordinates, we need to further transform them into variables that are also invariant under permutation of identical particles. These are obtained using invariant theory~\cite{Derksen2015-km,Braams2009-wi}. 
We generate invariant polynomials called primary and secondary invariants which are adapted to the permutational symmetry group on the rotationally invariant coordinates (distance-based or distance-angle), from which any polynomial that is invariant under permutation of like
atoms
can be expressed in a unique way. This gives 
\begin{equation}
    E_n^{\bZ}(\{\bR_i \}_{i=1}^n )
    = \sum_b s_{n,b}^\bZ P_{n,b}(\{p_{n,a}^\bZ\}),
    \label{eq:En-as-polynomial}
\end{equation}
where $\{p_{n,a}^\bZ\}_{a=1}^{n(n-1)/2}$ denote the primary invariants, 
$\{s_{n,b}^\bZ \}_b$ denote the secondary invariants, and $P_{n,b}$ are multivariate polynomials in the $n(n - 1)/2$ rotationally invariant coordinates $(\{u_{1j}\}, \{w_{i1j}\})$. 
These primary and secondary invariants are determined using the Computer Algebra System {\sc Magma}~\cite{Bosma1997-pq}. We refer to Refs.~\citenum{Braams2009-wi,Van_der_Oord2019-td} for further details of this construction.

The primary and secondary invariants depend on the symmetry groups which are induced by the element combinations and the rotationally invariant coordinates, but not directly on the identity of the elements. For example, the triplet of atoms with $\bZ = (1,1,6)$ and $\bZ = (1,6,6)$ have the same invariants because both contain one atom of one element and two atoms of a different element. 
We show the different possible invariants for body-orders 2 to 4 below that are used in this paper. They can also  be found e.g. in Bowman \& Braams~\cite{Braams2009-wi}.

\paragraph{Body-order 2.}
In this case, the only rotation-invariant coordinate is the distance separating the two particles, which is already permutation-invariant.
Therefore, a rotation and permutation invariant (RPI) representation of the two-body energy is
\[
    E_2^{(Z_1,Z_2)}(\bR_1,\bR_2) = E_2^{(Z_1,Z_2), {\rm RPI}}(u_{12}).
\]
In the notation of primary and secondary invariants this corresponds to choosing $p_{2,1} = u_{12}$, $s_{2,1} = 1$.
Note that the constant polynomial 1 is usually not considered as a secondary invariant, but we include it for convenience.

\paragraph{Body-order 3}
For three-body terms, there are three distinct element combinations possible, AAA, AAB and ABC, but only two cases have to be considered for distance-angle coordinates, since the center atom at which the angle is measured does not enter into the consideration of symmetry. We denote these two cases by $\star$AA, and $\star$AB, where $\star$ stands for the center atom.
We summarise a canonical choice (it is not unique) of invariants in Table~\ref{tbl:3b4b_invariants}.
Invariants for distance-based coordinates can be found in the Supplementary Information~S3. 

Note that considering the symmetry with respect to like atoms that are not the center atom of the environment, as is done in the case of distance-angle coordinates, leads to simpler invariants compared to the full symmetry as the considered symmetry group is smaller, but this is partly compensated by an additional sum needed to account for the different atom-centered environments in the computation of the energy.

\paragraph{Body-order 4}

For four-body terms, there are five distinct element combinations, which are AAAA, AAAB, AABB, AABC, ABCD. 
As for three-body components, only element combinations
$\star$AAA, $\star$AAB, $\star$ABC have to be considered for distance-angle coordinates, for which invariant polynomials are presented in Table~\ref{tbl:3b4b_invariants}.
Invariant polynomials for distance-based coordinates are presented in the Supplementary Information~S3.

\begin{table*}[]
    {%
    \begin{tabular}{r|c|c|c}
    & $\star$AAA & $\star$AAB    & $\star$ABC   \\
          \hline 
         $p_1$ & $u_{12} + u_{13} + u_{14}$ &
           $u_{14}$ & 
           $u_{12}$  \\
         $p_2$ & $w_{213} + w_{214} + w_{314}$
          &
          $u_{12}+u_{13}$
         & $u_{13}$  \\
        $p_3$ & $u_{12}^2 + u_{13}^2 + u_{14}^2$ &
         $w_{214}+w_{314}$ &  $u_{14}$  \\
         $p_4$ & $u_{12}w_{213} + u_{13}w_{314} + u_{14}w_{214}$ &
          $w_{312}$  
         & $w_{213}$  \\
         $p_5$ & $w_{213}^3 + w_{214}^3 + w_{314}^3$
          &
          $u_{12}^2+u_{13}^2$   & $w_{214}$ \\
         $p_6$& $u_{12}^3 + u_{13}^3 + u_{14}^3 + w_{213}^2w_{214} + w_{213}w_{314}^2 +  w_{214}^2w_{314}$
          &
           $w_{214}^2+w_{314}^2$   & $w_{314}$ \\
         \hline
         $s_1$ & 1 & 1 & 1 \\
         $s_2$ & $u_{12}w_{214} + u_{13}w_{213} + u_{14}w_{314}$ &
          $u_{12}w_{314}+u_{13}w_{214}$
            \\
          $s_3$ & $w_{213}^2 + w_{214}^2 + w_{314}^2$ & \multicolumn{2}{c}{
          \multirow{10}{*}{
              \begin{tabular}{r|c|c}
    & $\star$AA & $\star$AB \\
         \hline 
    $p_1$    &$u_{12} + u_{13}$ &  $u_{12}$\\ 
    $p_2$  &  $u_{12} u_{13}$ & $u_{13}$ \\ 
    $p_3$  &  $w_{213}$ &  $w_{213}$ \\ 
    \hline 
    $s_1$ & 1 &  1  
    \end{tabular}
          }
          }\\
          $s_4$ & $u_{12}^2u_{14} + u_{12}u_{13}^2 + u_{13}u_{14}^2$ \\
          $s_5$ & $u_{12}u_{13}w_{213} + u_{12}u_{14}w_{214} + u_{13}u_{14}w_{314}$ \\
          $s_6$ & $u_{12}w_{213}^2 + u_{13}w_{314}^2 + u_{14}w_{214}^2$ \\
          $s_7$ & $u_{12}^2w_{214} + u_{13}^2w_{213} + u_{14}^2w_{314}$ \\
          $s_8$ & $u_{12}w_{213}w_{214} + u_{13}w_{213}w_{314} + u_{14}w_{214}w_{314}$ \\
          $s_9$ & $w_{213}^2w_{314} + w_{213}w_{214}^2 + w_{214}w_{314}^2$ \\
          $s_{10}$ & $s_2 s_3$ \\
          $s_{11}$ & $s_2^2$ \\
          $s_{12}$ & $s_5^2$
    \end{tabular}
    }
\caption{\label{tbl:3b4b_invariants}3-body and 4-body primary and secondary invariants for distance-angle coordinates (with 3-body on the lower right). The $\star$ denotes the center atom, which also corresponds to index 1, and the subscripts follow the ordering of the atoms (e.g. $w_{213}$ is the angle between atom 2 and atom 3 measured at the central atom 1.)
}
\end{table*}

\subsubsection{Cutoffs and Basis}
For large molecules, we expect the contribution of terms that involve far-away atoms to be very small, hence we introduce a cutoff on the distance variables. This breaks the fundamental factorial scaling of the PIP scheme. The corresponding loss of accuracy depends on the application, can be observed numerically, and controlled by the cutoff.  Thus, for a given body-order component $\bZ = (Z_1,\ldots,Z_n)$, our final basis functions are given by
\begin{align}
  \label{eq:BkZ}
  B_{b{\bf k}}^{\bZ}(\RR)
  & =  \sum_{\substack{i_1 < \ldots < i_n\\ 
                    z_{i_l} = Z_l}} 
        F_{\rm cut}(\{\bR_{i_l}\}_{l=1}^n)   \\ 
  \notag 
  & \qquad \quad \times \left[ s_{n,b}^{\bZ} \prod_{a=1}^{n(n-1)/2}  (p_{n,a}^{\bZ})^{k_a}  \right],
\end{align}
where the invariants $s_{n,b}^{\bZ}, p_{n,a}^{\bZ}$ are evaluated at $\{\bR_{i_l}\}_{l=1,\ldots,n}$.
The tuples ${\bf k} = (k_a)_{a = 1}^{n(n-1)/2}$ are tuples of non-negative
integers with
\[
  {\rm deg}(s_{n,b}^{\bZ}) + \sum_a k_a {\rm deg}(p_{n,a}^{\bZ}) \leq D_{n},
\]
where $D_n\in\N$ is a prescribed maximum degree, and ${\rm deg}(s_{n,b}^{\bZ})$ and ${\rm deg}(p_{n,a}^{\bZ})$ are the total degrees of the primary and secondary invariants. We refer to Ref.~\citenum{Van_der_Oord2019-td} for further discussion of these basis functions.

To specify $F_{\rm cut}$ we choose a univariate cutoff function $f_{\rm cut}(r)$ which is smooth and vanishes outside some cutoff radius
$r_{\rm cut}$, and then define 
\[
   F_{\rm cut}(\{\bR_{i_l}\}_{l=1,\ldots,n}) 
   =
     \prod_{j = 2}^n f_{\rm cut}(r_{1j})%
\]
In practice, we choose a cutoff radius $r_{\rm cut}$ and a cutoff parameter $r_{\rm cut}'<r_{\rm cut}$, and we use
\begin{equation}
    f_{\rm cut}(r)
    = 
    \begin{cases}
     1 & 0 \le r < r_{\rm cut}' \\ 
     \frac{1}{2} {\scriptstyle \left( \cos\left(\pi \frac{r-r_{\rm cut}'}{r_{\rm cut}-r_{\rm cut}'}\right) \! + \! 1\right) }
     & r_{\rm cut}' \le r \le r_{\rm cut} \\
     0 & r > r_{\rm cut}.
    \end{cases}
    \label{eq:fcut}
\end{equation}
In the assembly of the total potential energy,
only clusters respecting the cutoff condition $F_{\rm
  cut}(\{\bR_{i_l}\}_{l=1,\ldots,n})>0$ are taken into account.

\subsubsection{Summary of the basis generation}

Finally, each term in the total energy expression~\eqref{eq:bo-expansion} is expanded as a linear combination of the basis functions defined in \eqref{eq:BkZ} and we are left with the determination of the coefficients $(c_{b{\bf k}}^{\bZ})$  
which will be described in Section~\ref{sec:LSQ_fitting}.
For now, let us summarize the generation of the symmetry-adapted polynomial basis.
First, we choose the body-order components taken into account in the expansion~\eqref{eq:bo-expansion}, indexed by $\bZ$. In practice, we will very often choose to take all possible components up to body-order 4, that is 12 components for systems with two chemical elements A and B (AA, AB, BB, AAA, AAB, ABB, BBB, AAAA, AAAB, AABB, ABBB, BBBB).%
Then, for each component,%
\begin{enumerate}
    \item we choose rotationally invariant coordinates, which are either distance-based or distance and angles based.
    \item we compute the primary and secondary invariants relative to the corresponding permutation group using {\sc Magma}~\cite{Bosma1997-pq}.
    \item we choose a cutoff function and a cutoff radius.
    \item we choose a maximum polynomial degree and consider all possible basis functions with a lower degree.
\end{enumerate}
The total energy is then expanded as a linear combination of these basis functions, as
\begin{align}
    \E(\RR) &= \sum_n \sum_{\substack{\bZ \text{ s.t.}  \\ \#\bZ = n}}
    \sum_{b, {\bf k}} c_{b{\bf k}}^{\bZ} 
    B_{b{\bf k}}^{\bZ}(\RR). \label{eq:lin_comb}
\end{align}

\subsection{Least-squares fitting}
\label{sec:LSQ_fitting}

It remains now to determine the coefficients $(c_{b{\bf k}}^{\bZ})$ in the linear expansion \eqref{eq:lin_comb}. For this, we solve a linear least squares problem, where the training set is composed of atomic configurations $\RR$ with their corresponding energy $\Ee_{\RR}$ and forces $\Fe_{\RR}$.
The minimized functional is of the form
\begin{equation} \label{eq:lsq_fcnl}
\begin{split}
   J & =
   \sum_{\RR} \Big( \;
      \left(\frac{W_{E}}{N_{\rm at}}\right)^2 \Big| \E(\RR) - \Ee_{\RR} \Big|^2  \\
      & \qquad \quad + W_{F}^2 \left| F(\RR) - \Fe_{\RR} \right|^2 \Big) + \text{ Reg },
\end{split}
\end{equation}
where $W_E$, $W_F$ are weights that may depend on the configurations
$\RR$, $N_{\rm at}$ is the number of atoms in the system, $F(\RR)$ are the forces
computed from the energy functional $\E(\RR)$, and Reg contains all the regularisation terms that will be described in the next section.

Without regularisation terms, $J$ is quadratic in the unknown polynomial coefficients $c_{b{\bf k}}^{\bZ}$, hence minimizing $J$ can be done by solving a 
standard linear least-squares problem
\begin{equation}
   \label{eq:LSQ_system}
   \min_{\bf c} \|A{\bf c} - Y\|_2^2,
\end{equation}
which we solve using a QR factorisation. We will show below that adding regularisation terms does not change the linear structure of the problem.

\subsection{Regularisation}

In order to improve the smoothness of the potential as well as its extrapolation capabilities, we use two regularisation techniques described in Ref. \citenum{Van_der_Oord2019-td}. 

First, we use a Laplace regulariser, which adds a contribution to the least-squares functional for each body-order component of the form 
\begin{equation}
   \label{eq:reg_sumfcnl_laplace}
   J_n^\bZ = \frac{\gamma_n^\bZ}{J} \sum_{j = 1}^J
   \Big| \Delta \big[E_n^\bZ({\bf u}_j)\big]\Big|^2,
\end{equation}
where $\gamma_n^\bZ$ is an adjustable regularisation parameter, and the second derivatives of $E_n^\bZ$ are approximated with finite-difference. The points $({\bf u}_j)$ are chosen according to a Sobol sequence.
This regularisation penalises large variations of the potential, hence promotes the smoothness of the potential. Varying the paramaters $\gamma_n^\bZ$ allow  balancing between the accuracy of the fit and the smoothness of the potential.

\begin{figure}[ht]
    \centering
    \includegraphics[width=8.0cm]{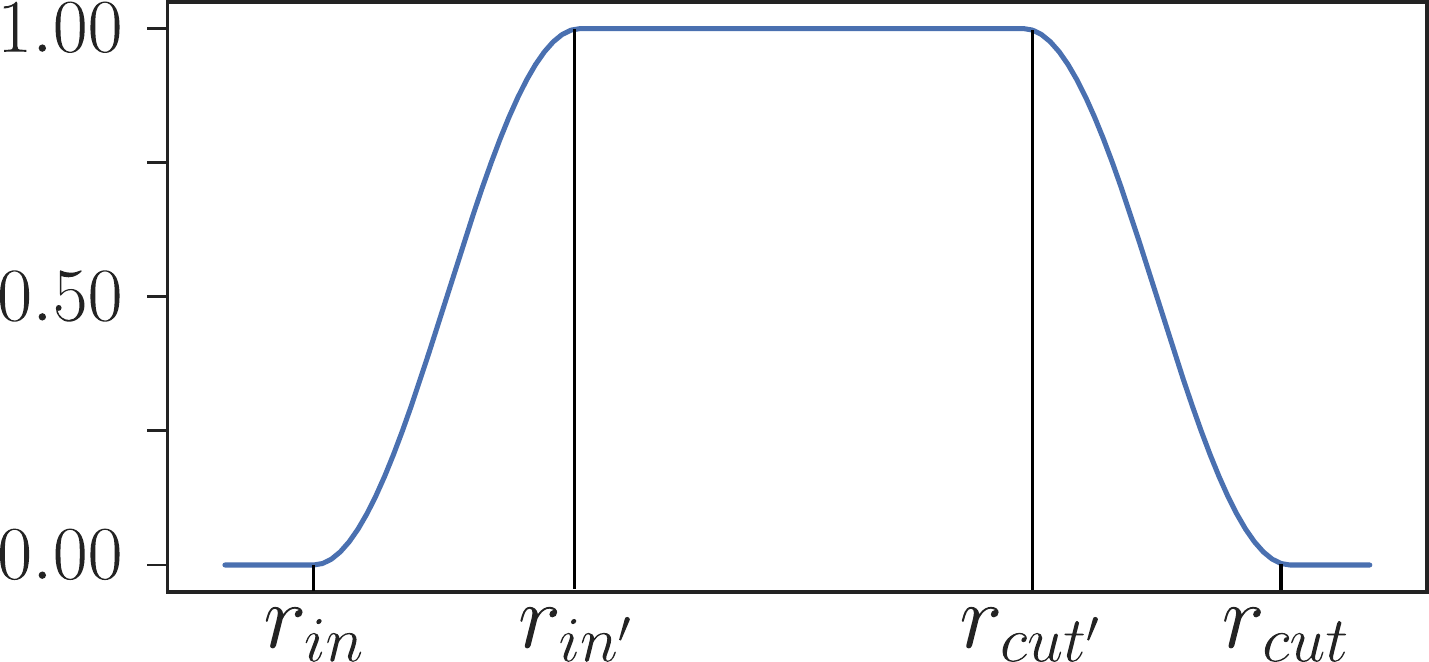}
    \caption{ \label{fig:2s_cutoff}Two-sided cutoff function $f_{\rm cut}^{2s}$ with cutoff radii $r_{\rm cut},r_{\rm in}$, and parameters
$r_{\rm cut}',r_{\rm in}'$. }
   
\end{figure}

Second, we use a two-sided cutoff for 3B and 4B components, and a simple analytic repulsive 2B function for small interatomic distances, in order to prevent ``holes'' in the PES coming from polynomial oscillations in this region~\cite{Nandi2019-bl}. 
The two-sided cutoff consists in replacing the cutoff functions $f_{\rm cut}(r)$ by a function which satisfies $f_{\rm cut} = 0$ on both $[0,
  r_{\rm in}]$ and $[r_{\rm cut}, \infty)$, e.g.
\[
f_{\rm cut}^{2s}(r) = (1 - f_{\rm cut,in}(r)) f_{\rm cut,out}(r),
\]
where $f_{\rm cut,out}(r),f_{\rm cut,in}(r)$ are the cutoff functions defined in \eqref{eq:fcut} respectively with cutoff radii $r_{\rm cut},r_{\rm in}$, and parameters 
$r_{\rm cut}',r_{\rm in}'$, as shown on Figure~\ref{fig:2s_cutoff}.

For the two-body components, we start by solving the linear least-squares problem with the two-body components defined on the whole interval $[0,r_{\rm cut})$. We choose a splining point $r_{\rm S}$ that is sufficiently small so as to not influence the training set error, specific values are given in the following section.  
The new two-body components with repulsive core are defined such that
\begin{equation} \label{eq:repulsive_core}
\begin{split}
   \tilde{E}_2(r) &:= \begin{cases} E_2(r), & r \geq r_{\rm S},
     \\ E_{\rm rep}(r), & r < r_{\rm S},
   \end{cases} \\
   E_{\rm rep}(r) &= e_\infty + \beta r^{-1} e^{- \alpha r},
\end{split}
\end{equation}
where $e_\infty < E_2(r_{\rm S})$ is a parameter adjusting the steepness of the potential, and $\alpha, \beta$
are chosen such that $\tilde{E}_2$ is continuous and continuously
differentiable at $r_{\rm S}$. 

Note that the regularised least-squares problem is still linear, hence the regularisation procedure does not affect the computational cost of the fit.

\section{Methods}
\subsection{Fits and hyper-parameters}
We now describe the details of the fitting process and training data generation to construct aPIPs for molecules. As detailed in Section~\ref{sec:results}, we explore the use of aPIPs for fitting the PES of individual molecules (trained and tested independently from one another), as well as for fitting a combined force field. The combined force field is fit to data from multiple linear hydrocarbons, and is then shown to be  accurate for both the molecules it has been fit to as well as to slightly longer linear hydrocarbons.  We used the same hyper-parameters in all fits, as shown in Table~\ref{table:Fitting_params_pips}, except for the individual N-methylacetamide PES which was fit with a 4B maximum degree of 6 and the combined force field which used a cutoff of $r_{\rm cut}'$ = 2.75 and $r_{\rm cut}$ = 3.25\AA. For N-methylacetamide, as there are four different elements present, the number of basis functions is very large when the maximum degree for the 4B term is 10. Therefore, a lower maximum degree was used in this case to reduce the number of basis functions and still allow a potential to be fit for NMA with only 1000 structures. When performing individual molecule fits, the performance of the potential is relatively insensitive to the choice of outer cutoff. However, for the combined fit reducing the outer cutoff resulted in improved performance on the testing set. 

When fitting individual molecular PESs, the hyperparameters could be optimised anew for each molecule and this would no doubt improve accuracy, but we are more interested in using a generic parameter set which does not need to be tuned and greatly reduces the effort required to create new fits. The fact that the resulting PESs are still very accurate suggests that the potentials could be made transferable across molecules.
The potentials created include all possible 2B, 3B, and 4B terms; e.g. for butane, the 4B terms are HHHH, CHHH, CCHH, CCCH, CCCC. Not including some of these would result in a smaller number of basis functions and in some cases may not influence the accuracy of the potentials, for example one might surmise that the HHHH terms are not necessary. However, we have included all terms in our fits in order to eliminate the necessity of such manual choices.

The number of basis functions depends on the number of different elements, the maximum body order and polynomial degree. By way of example, the alkane fits in this work (beyond butane) use 13629 basis functions. 

\begin{table}
\begin{tabular}{l|c|c}
Parameter                        & Symbol         & Value      \\ \hline
Max degree-2B  & $D_2$                      & 12                              \\ 
Max degree-3B                    & $D_3$  & 10                              \\ 
Max degree-4B                    & $D_4$  & 10                              \\ 
Outer Cutoff                   & $r_{\rm cut}', r_{\rm cut}$  & 4.00, 5.50  \AA                   \\
{Inner Cutoff}                    & $r_{\rm in}, r_{\rm in}' $ & 0.70, 0.80  \AA                   \\
{Weight-Ratio}                      & $W_E:W_F$  & 100:1           \\ 
Regularisation & $\gamma_n^{\bZ}$ & 0.05                            \\ 
Radial transform                     &  $u$  & $e^{( - 2.5(r/1.54 - 1) )}$ \\ 
\end{tabular}
\vskip 0.5cm
$r_{\rm S}$  (\AA)
\begin{tabular}{lllll}
 {CC: 0.93}&{CH: 0.56} & {HH: 0.27} & {CO: 0.98} \\
 {OH: 0.61} & {NO: 1.02 } & {NC: 0.96} & {NH: 0.59} \\
 {OO: 1.04}& {NN: 0.99}&&
\end{tabular}
\caption{\label{table:Fitting_params_pips}The hyperparameters used for fitting the aPIPs. The $r_{\rm S}$ values that define the switch-over to a repulsive core, $E_{\rm rep}$, in \eqref{eq:repulsive_core}, correspond roughly to the point where the ZBL potential is 20~eV. }

\end{table}

\subsection{MD Training Data}
The majority of the training data in our fits is obtained by taking snapshots from MD trajectories with a temperature of 1500K. %
To reduce the computational cost, these MD simulations were performed using DFTB~\cite{Elstner1998} with the \hbox{mio--1--1} parameter set, using the NVT ensemble with a Langevin thermostat and a friction coefficient of 0.002. The time step was 0.1~fs and samples were taken 1000 time steps apart. 
A total of 1000 structures were collected for each molecule, except for ethanol and N-methylacetamide, where the temperature was reduced to 800~K to prevent bond dissociation. We emphasise that with a fixed sized basis set, we expect {\em convergence} of the coefficients and thus a large number of data points were collected to ensure that. The evaluation cost of the aPIPs force field is independent of the number of training data points. We made no attempt to study the minimum size and composition of the optimal training set, this is left for future work. Energy and force data for the force field fit was then obtained by reevaluating the snapshots from the DFTB-MD using Molpro~\cite{MOLPRO} with a B3LYP hybrid DFT functional~\cite{Stephens1994, Lee1998, Becke1993} and 6-31G** basis set.

\subsection{Additional Training Data}
In addition to the high temperature MD, there are two more sources of training data that take account of the special structure of PESs. The first
issue is that the repulsive potential~\eqref{eq:repulsive_core} we add at small distances is not designed to accurately reproduce the potential energy, rather it is kept simple in order to ensure that it is repulsive and does not introduce additional local minima. 
This means that the splining point $r_S$ at which it is turned on is chosen well below the distance which we expect to encounter between atoms even at the high temperature of 1500K. 
The smooth transition to the repulsive potential is thus aided by manually adding dimer configurations of each pair of elements to the data set. 
One choice would be to use the same level of quantum mechanics for these as for the rest of the data set, but we opt instead to use the ZBL functional~\cite{Ziegler1985-sd}, which is fit to Hartree--Fock data and has better accuracy than DFT at small distances. 
This ZBL set consists of 55 dimers with interatomic distances in the range of [0.1~\AA,5.50~\AA]. 
In the least squares fit, we reduce the weight if the ZBL set in a ratio of 100:1 relative to the rest of the training set, so that the ZBL data does not noticeably influence the accuracy of the fit in regions of configuration space where other data exists.

The second additional training data concerns only the molecules butene and ethene. The HCCH dihedral around the double bond in these molecules will not be sampled during the MD simulations as the energy barrier is too high. Therefore, the HCCH dihedral energy scans around the double bond are added to the training set, with 12 data points in each scan. The weight of this dihedral scan data is increased in the ratio 2:1, to take into account of the small number of samples in this additional data set. %

\subsection{Iterative Fitting Process}
\label{sec:iterative}
One of the key goals in using aPIPs is that ``holes'' in the potential should be eliminated. We consider any region of configuration space a hole which has lower  energy than the energy of the molecule's locally optimized structure when starting from the true equilibrium geometry. %
The presence of holes in PIP fits are well documented~\cite{Nandi2019-bl}, and many  overparametrised ML models are also in danger of having holes due to the high dimensionality of the molecular representation they use. 
In order to try and eliminate holes, we introduce an iterative fitting process that systematically searches for holes in configuration space. Various iterative fitting methods exists, with either geometry optimisation~\cite{Deringer:2017ea,Bernstein:2019jw}, MD or Diffusion Monte Carlo (DMC)~\cite{Debiec2016, Nandi2019-bl} used to sample the PES. Alternatively, active learning can be used, where the uncertainty of the prediction for a set of test structures is quantified and the most uncertain structures are added to the training set~\cite{Gubaev:2019gz,Smith2019}. The dimensionality of our energy terms is much smaller than in the cases of the cited works, and so we opt for a different strategy. Using the Sobol quasirandom sequence~\cite{Niederreiter-Sobol}, we create a ``Sobol test set'' for each molecule according to the following procedure: 
\begin{itemize}
    \item The optimized structure of the molecule is calculated with the B3LYP hybrid functional and 6-31G** basis set, and the geometry is converted to internal coordinates. 
    \item A Sobol sequence of length 100,000 is then produced with a dimension equal to the number of internal coordinates. 
    \item The elements in the sequence correspond to the displacement of the internal coordinates from their equilibrium position, scaled such that the range of bond lengths, angles and dihedral displacements for non-cyclic molecules is $\pm0.30$~\AA, $\pm10^{\circ}$ and $\pm10^{\circ}$ respectively, for cyclic molecules these ranges are halved. Rotatable dihedrals are identified and allowed to take a value between 0-360$^{\circ}$. 
    \item The scaled Sobol sequence displacement vectors are then used to generate the set of 100,000 ``Sobol test set'' structures.  
    \item Finally, we check for clashing atoms: any structure with atoms closer than the corresponding $r_S$ (see Table~\ref{table:Fitting_params_pips}) is removed from the set.  
\end{itemize}
The resulting Sobol test set has a diverse range of structures that sample the space around the equilibrium position more uniformly than if we did stochastic sampling. It is very fast to generate and does not rely on carrying out MD or DMC with either DFT or even a preliminary force field. Doing the latter could lead to poor results because the trajectory can get trapped in a hole, or result in bond dissociation. 

\begin{figure}[ht]
    \centering
    \includegraphics[width=3.0in]{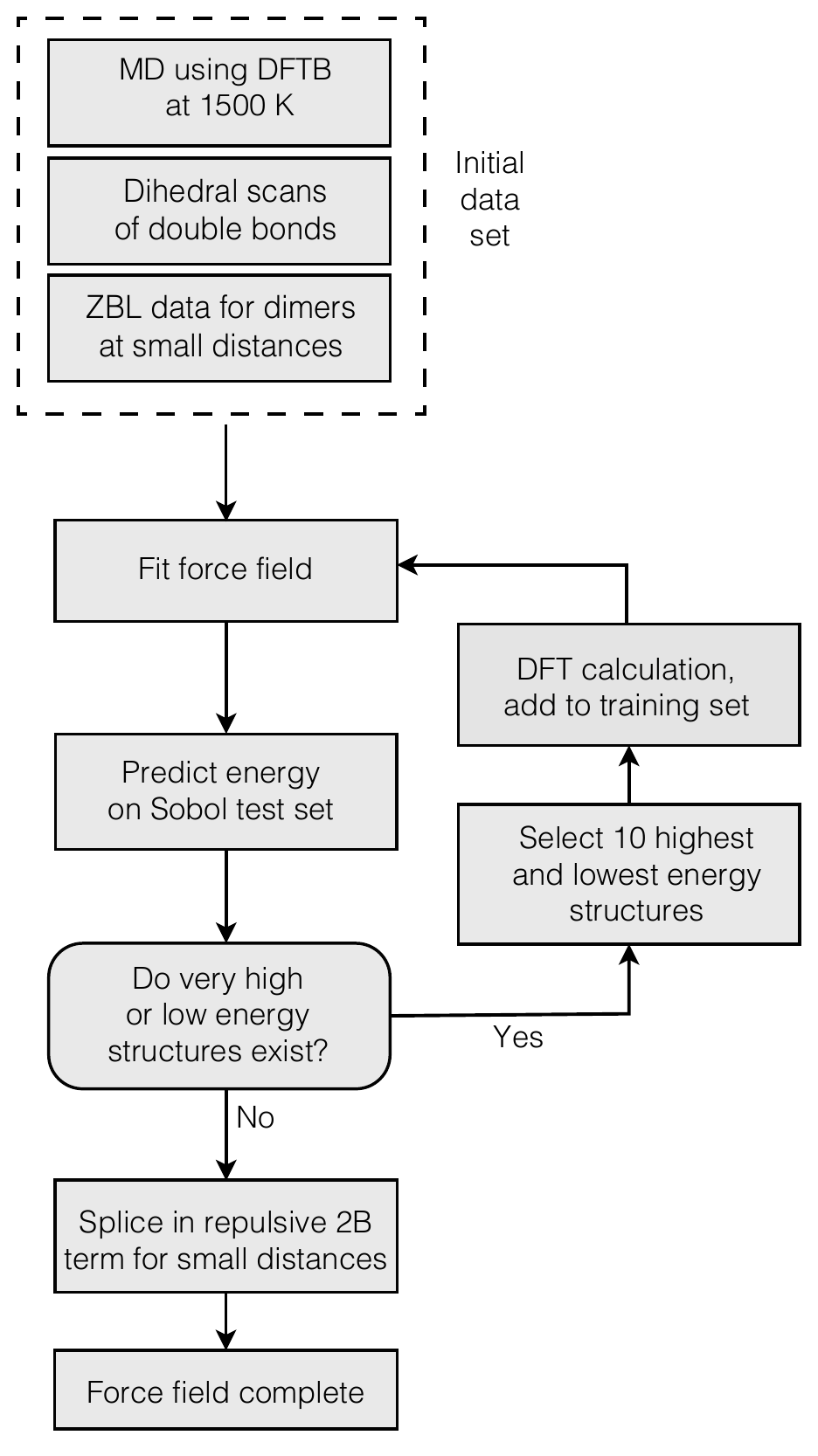}
    \caption{ \label{fig:flowchart_fitting}The flowchart shows the stages we used to build the database and fit the aPIP force field.}
   
\end{figure}

The entire fitting process is summarized in Figure~\ref{fig:flowchart_fitting}. In each iteration, the potential energy of the Sobol test set is calculated using the current force field. DFT calculations are performed for the five lowest energy structures, and up to five highest energy structures (with energies above 2.5~eV/atom), and added to the training set in the next iteration. This process is repeated until there are no structures with an energy below that of the equilibrium structure. 

\subsection{Empirical Force Field Comparison}
\label{sec:classical_ff_method}

To demonstrate the improved performance of aPIPs over an empirical force field, we parametrised a simple force field for methane on the same training set. The maximum body order is five, but the functional form is very limited and the number of free parameters is very small. The precise form for the empirical force field is given by,
\begin{equation}
\small
\begin{split}
     &E =\sum_{bond} [K_{b2}(b-b_0)^2 + K_{b3}(b-b_0)^3 + K_{b4}(b-b_0)^4 ] \\
    &+ \sum_{angle}  [K_{a2}(\theta-\theta_0)^2 + K_{a3}(\theta-\theta_0)^3 + K_{a4}(\theta-\theta_0)^4 ] \\
    &+ \sum_{bond/bond}K_{bb}(b-b_0)(b'-b'_0) \\
    &+ \sum_{bond/angle}K_{ba}(b-b_0)(\theta-\theta_0) \\
    &+ \sum_{angle/angle}K_{aa}|(\theta-\theta_0)(\theta'-\theta'_0)| \\
\end{split}
\end{equation}
The parameters were determined by minimizing the functional given in Equation~\eqref{eq:lsq_fcnl}. The $W_E$ and $W_F$ terms in Equation~\eqref{eq:lsq_fcnl} were the same as those used for the aPIP potential. For simplicity and given the simple functional functional form and large amount of data used, no regularization was needed in this case.

\section{Results}
\label{sec:results}

\subsection{Convergence: tests on methane}
\label{sec:convergence_test}

\begin{figure}[ht]
    \centering
    \includegraphics[width=3.2in]{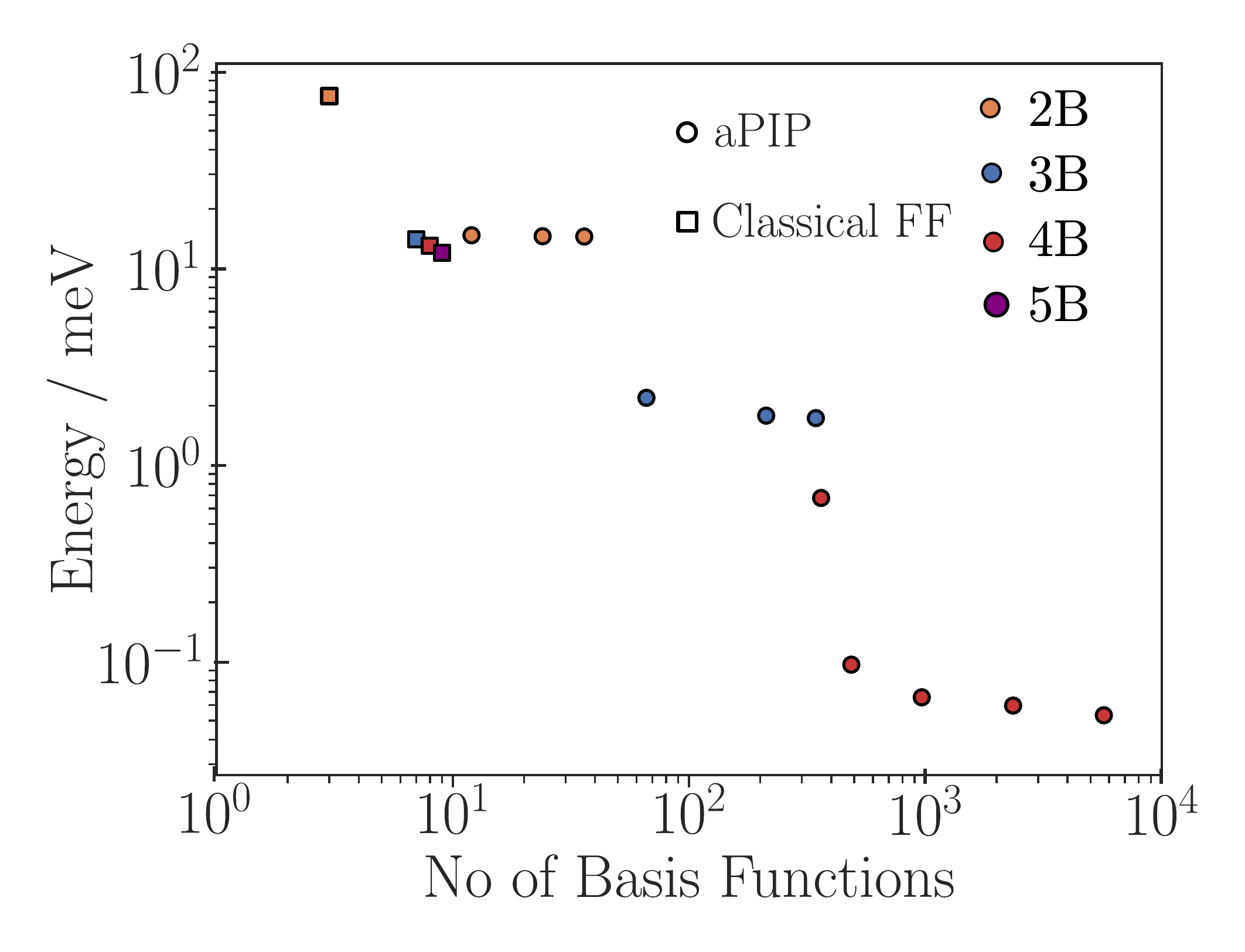} 
    \caption{\label{fig:methane_no_basis_functions}The change in the training set RMSE with the number of basis functions and body order for the methane aPIP potential (circles), up to body-order 4.
    Also shown are the errors of an empirical potential fit as the corresponding bond, angle etc terms are added (squares), up to body-order 5. 
    }
    
\end{figure}

Before we discuss the aPIP force field's performance for a set of small molecules, we first examine its convergence properties, tradeoff between speed and accuracy as controlled by the basis set size, and the effects of regularisation and the iterative fitting process, all for the case of the individual fit to methane.  Figure \ref{fig:methane_no_basis_functions} shows the decrease in the energy root mean square error (RMSE) with the increasing number of basis functions and body order. The number of basis functions is dependent both on the maximum polynomial degree and body order, and for a fixed body order, the RMSE levels off to the minimum error possible for that body order. The minimum error for 3B is about $2\times10^{-3}$ eV/per atom, and increasing the body order to 4B without changing the overall number of basis functions results in a big drop in the RMSE. In contrast, this is not the case for the empirical force field, for which there is very little gain beyond adding 3B terms. 

In the Supplementary Information, we show that the distance-based coordinates are less accurate than the distance-angle coordinate system. %
The decision to use distance-angle coordinates for our force field is further strengthened by the results shown in S2, with the normal mode recreation for butene significantly better. The convergence of additional properties with the number of basis functions is discussed further in S2.

\begin{figure*}[]
    \centering
    \includegraphics[width=6.5in]{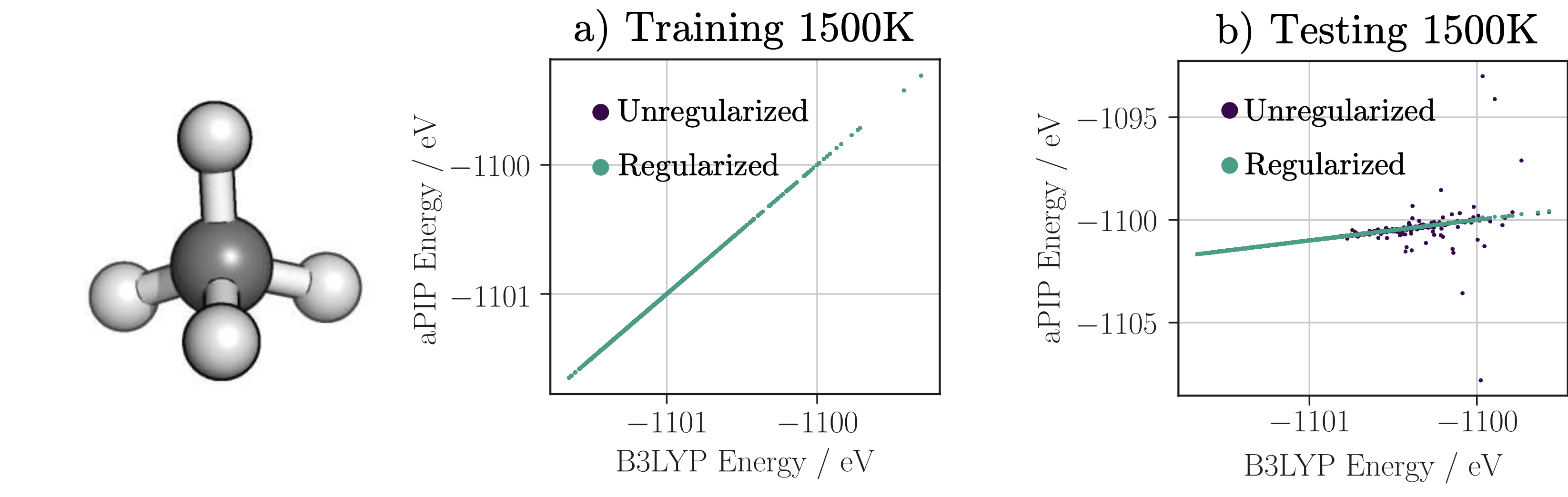}
    \caption{ \label{fig:methane_reg_no_reg}Comparison of target and predicted energies for the training and test sets, for both regularised and unregularised fits. 
    }
   
\end{figure*}

Although, as stated in the introduction, the aPIP formulation is the bridge between the low body order empirical force fields and high dimensional ML potentials, from the point of view of number of degrees of freedom we are interested in the regime when it is closer to the latter. Having thousands of degrees of freedom means that regularising the least squares fit is necessary in order to avoid overfitting. The regularisation effect is shown in Fig.~\ref{fig:methane_reg_no_reg}. For the unregularised potential, the test set RMSE is about a 100 times larger than the training set RMSE. 

\begin{figure}[]
    \centering
    \includegraphics[width=3.2in]{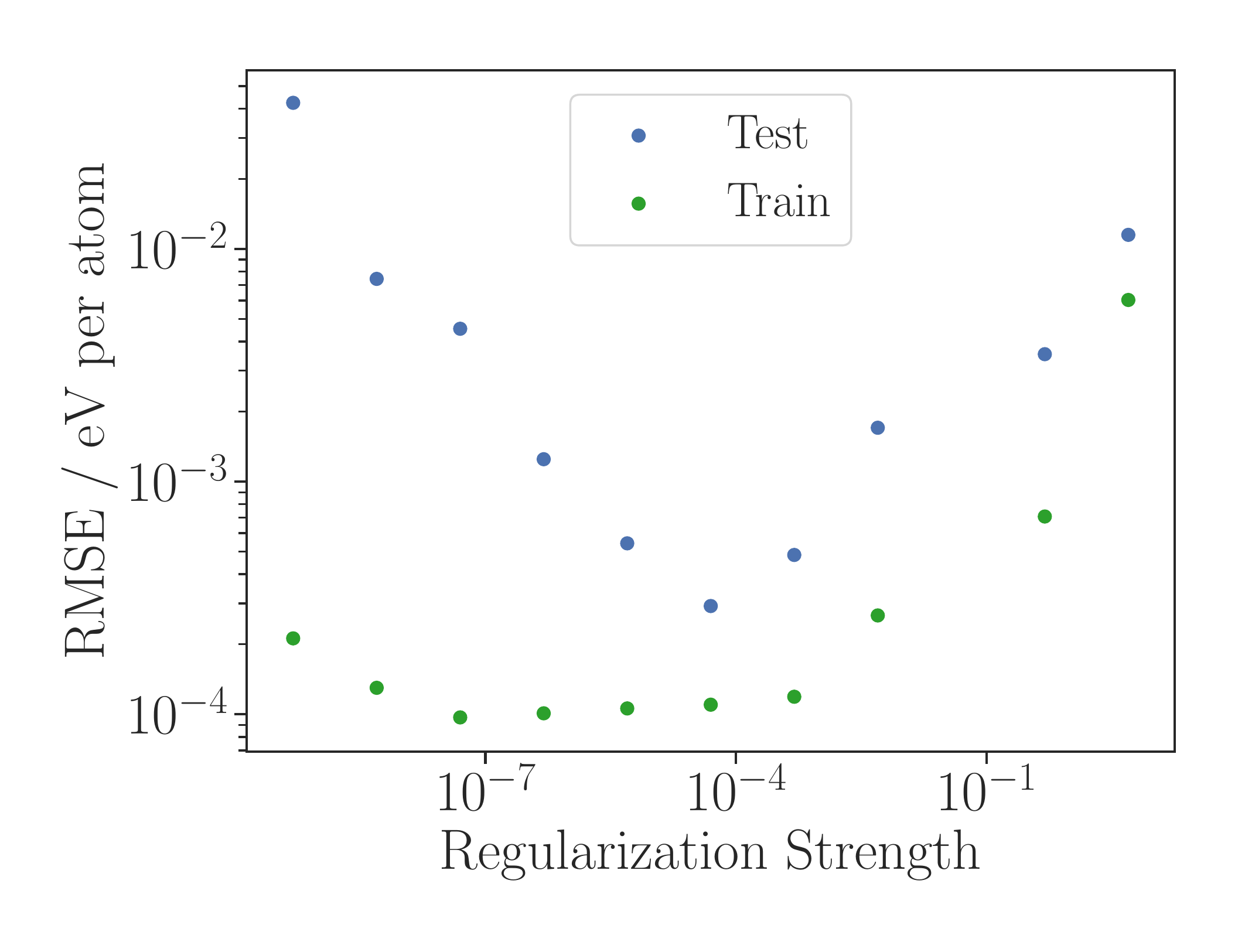}
    \caption{\label{fig:methane_reg_rmse}The energy RMSE of methane on the 1500K training and test set (8000 structures) as a function of regularization strength ($\gamma_n^{\bZ}$).}
    
\end{figure}

Figure~\ref{fig:methane_reg_rmse} shows the effect of changing the regularisation strength $\gamma$ on the RMSE on the training and test sets. The training set is as described in the previous section, ZBL data and the iterative structures for the regularized fit are included in all potentials, whereas the test set is composed of 8000 independently sampled structures from the same 1500K MD simulation. For methane, the optimal regularization strength is at approximately 10$^{-4}$. The training set RMSE is relatively constant up until $10^{-2}$ when an increase starts. Given that we opted to use a single value for fitting all molecules in this paper, the higher regularisation strength of 0.05 is used, because for larger molecules that is beneficial due to the larger number of basis functions.  

\begin{figure}[]
    \centering
    \includegraphics[width=3.2in]{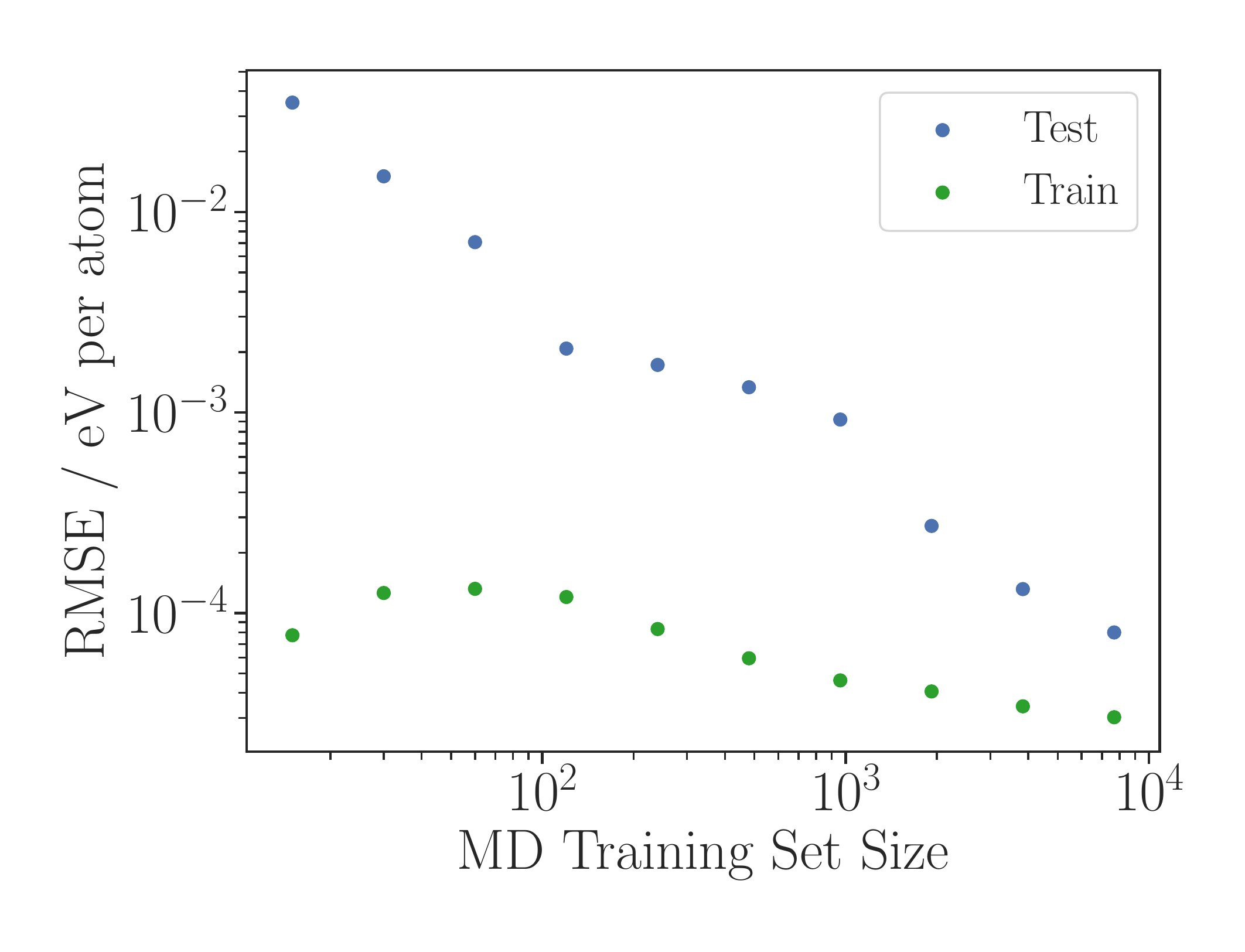}
    \caption{\label{fig:methane_increasing_data}Energy RMSE of methane on a the 1500K training and test set (1000 structures) as a function of the MD sampled training set size. All other parameters are as in Table~\ref{table:Fitting_params_pips}. Note that the errors at 1000 training structures are slightly higher than in Table~\ref{table:rmse_test_train} because the the potentials reported there include additional training data. 
    }
    
\end{figure}

Finally, Fig.~\ref{fig:methane_increasing_data} shows the RMSE as a function of the number of configurations in the MD training set (with no additional data), with the other parameters as in Table~\ref{table:Fitting_params_pips}. The decrease in RMSE on a fixed test set of 1000 independent structures shows no sign of saturation, and reaches $1\times10^{-4}$ for 8000 training configurations. We kept the regularisation strength fixed here, but if it were optimised separately for each training set size, the error would go down further and the difference between the errors on the training and test set be also naturally further reduced. 

\subsection{Iterative Fitting}
\begin{figure}[h]
    \centering
    \includegraphics[width=3.2in]{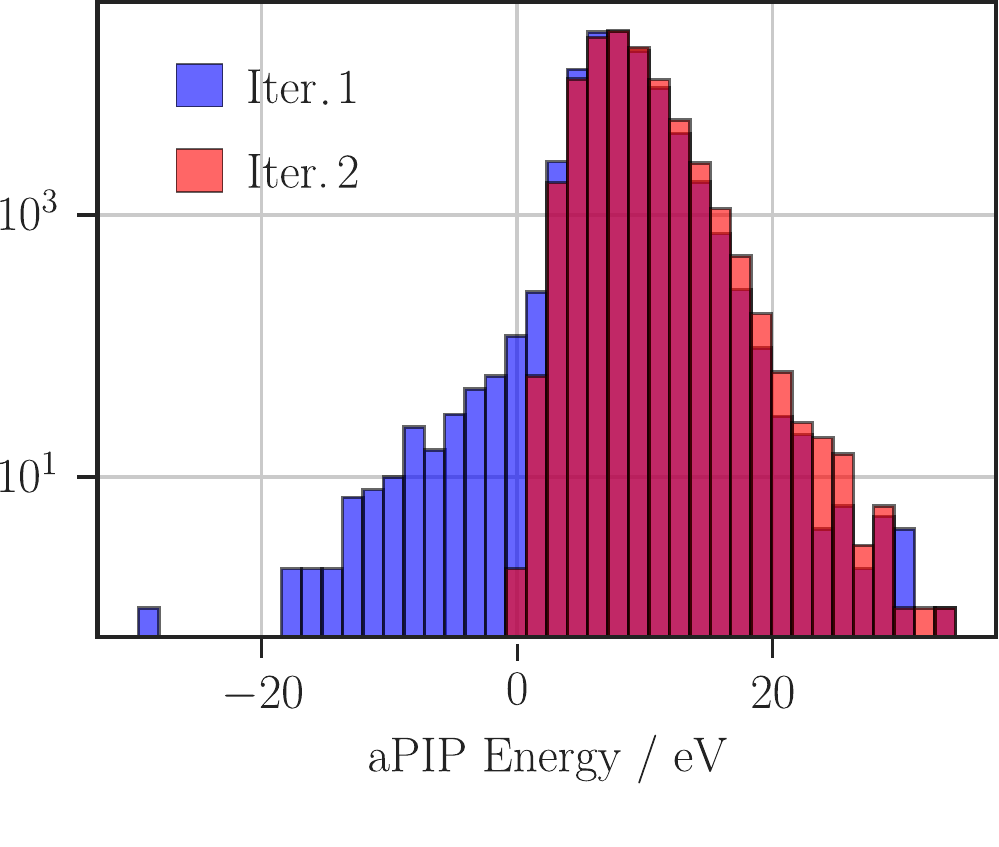}
    \caption{\label{fig:butane_hist_iter}The energy distribution of the Sobol test set for the butane aPIP potential at iteration 1, and iteration 2 (trained with five additional Sobol test set structures). The energy is shifted so that true geometry optimised butane molecule is at 0.0~eV.}
    
\end{figure}

In Section \ref{sec:iterative} we introduced an iterative fitting algorithm that samples structures derived using the Sobol quasirandom sequence. Figure~\ref{fig:butane_hist_iter} demonstrated the effect of a single iteration on the energy distribution on the Sobol test set. The original fit (labelled ``Iter 1'' in the figure) resulted in thousands of structures having a lower energy than the true equilibrium structure. The addition of just five of the lowest energy structures results in a considerable change in the energy distribution seen at the second iteration and all the structures with unrealistically low predicted energies disappear. 

\subsection{Individual Molecule Force Fields}
\label{sec:results_indiv_molecule}

In this section, the performance of the aPIP force fields with up to 4-body terms for individual molecule PESs is evaluated for 14 small organic molecules. A combined fit to alkanes is discussed in the next section. Although ultimately we are interested in general force fields, considering the individual fits is interesting for a number of reasons. It enables comparison to other models that also target PESs one at a time (such as the PIP scheme and sGDML). Characterising the lowest possible error within a given body order and polynomial degree for individual fits is helpful when thinking about the errors of a general force field because it informs us of the extent to which the combined fit is forced to make compromises between fitting to data corresponding to different molecules. 

\subsubsection{RMSE for Training and Testing Set}

A most basic test of the performance of a force field is the RMSE of the energies for a training and testing set. Table \ref{table:rmse_test_train} summarises the energy RMSE for the 14 molecules tested, further graphs are given in the Supplementary Information S1. We show total energy errors, and also error/atom, because as molecules get larger, we expect (when keeping the training set size constant) that the total energy error would go up, but the error/atom stay bounded. The table shows that this is mostly true, the test set error/atom stays near or below 3~meV/atom for molecules with only single bonds, and below 5~meV/atom for molecules with double bonds.

\begin{table}[h!]
\begin{tabular}{rr|r|r|r}
&&\multicolumn{3}{c}{Energy RMSE (meV)} \\ \cline{3-5} 
Molecule   & Atoms & \makecell{Train\\1500K*} & \makecell{Test\\300K} & \makecell{Test\\1500K*} \\ \hline
\multicolumn{5}{l}{\textbf{Regularised 4-body aPIP}}\\
\multicolumn{5}{l}{(individual molecule fits)}\\
Methane    & 5     &0.3 &  0.2 &  1.6 (0.3)\\
Ethane     & 8     &1.8 &  1.2 &  7.8 (1.0)\\
Propane    & 11    &3.0 &  2.9 & 12.7 (1.2)\\
Butane     & 14    &8.2 &  7.0 & 29.1 (2.1)\\
Pentane    & 17    &12.8 & 13.0 & 50.1 (2.9)\\
Hexane     & 20    &19.6 & 28.1 & 65.8 (3.3)\\
Adamantane & 26    &11.5 &  4.1 & 22.0 (0.8)\\
Ethene     & 6     & 3.0 &  2.1 & 21.9 (3.7)\\
Butene     & 12    &13.3 & 16.0 & 56.9 (4.7)\\
Butadiene  & 10    & 7.9 &  5.2 & 31.5 (3.2)\\
Benzene    & 12    & 4.3 &  1.9 &  8.8 (0.7)\\
Methylbenzene & 15 & 8.5 &  4.1 & 25.7 (1.7)\\
Ethanol    & 9     & *3.3 &  1.4 &  *5.6 (0.6)\\
NMA        & 12    & *4.0 &  1.8 &  *4.6 (0.4)\\ \hline
{\em mean} &       & 7.2 &  6.3 & 24.6 (2.2)\\
\\
\multicolumn{5}{l}{\textbf{Unregularised 4-body aPIP}}\\
\multicolumn{5}{l}{(individual molecule fits)}\\
{\em median}&      & 4.4  &  3.4  & 211 \\
{\em mean} &       & 5.3&  $10^6$ & $10^8$\\
{\em maximum}&     & 16   & $10^7$ & $10^9$\\
\hline
\\
\multicolumn{5}{l}{\textbf{Combined 4-body aPIP fit}}\\
Methane & 5 & 4.38 & 0.98 & 3.47 (0.7)\\ 
Ethane  & 8 & 8.27 & 2.85 & 12.25 (1.5)\\
Propane & 11& 17.58 & 8.19 & 22.35 (2.0)\\
Butane  & 14& 23.78 & 10.54 & 30.65 (2.2)\\
Pentane & 17& 26.15 & 16.94 & 44.27 (2.6)\\
Hexane  & 20& 31.16 & 24.36 & 65.90 (3.3)\\
Heptane & 23& -  & 35.06 & 118.36 (5.1)\\
Octane  & 26& - & 44.84 & 156.62 (6.0)\\ 
\end{tabular}
\caption{\label{table:rmse_test_train}
The RMSE of the energies for training and test sets is given for the fourteen molecules tested. The higher temperature testing set is taken from 1500K MD for all molecules, except ethanol and NMA (marked by *) where it is 800K, this is due to bond dissociation occurring for the higher temperature. Energy errors per atom are shown in parentheses. 
}
\end{table}

Table~\ref{table:rmse_test_train} also shows that the 300K test set RMSE is comparable to the training set error. The configurations sampled by the 300K MD will be well within the sample of structures that the potential is fit to and are therefore well reproduced by the aPIP potential. The error for the higher temperature MD test set is several times higher than the training error. This is because structures that are not well represented by the training data will be present in the higher temperature MD. As discussed in Section \ref{sec:convergence_test}, an increase in the number of structures in the training set will result in a decrease in the test set error (and this was demonstrated for methane). With a fixed body order and basis set size, there is of course a saturation to a minimum error. As an example, when a fit is made to butane with the same parameters but 10,000 training structures instead of 1000, the energy RMSE for the 1500K test set falls by 20\% to 22.7~meV, a substantial decrease, but not as large as that for methane. There is no expectation that the rate of convergence is the same for different sized molecules. A detailed study of convergence rates for different molecules is left for future work.

\subsubsection{Comparison to Empirical Force Fields}
In order to directly assess the enhanced accuracy of the aPIP model over
empirical force fields of the same body order, we parametrised one for methane as described in Section \ref{sec:classical_ff_method}. The key differences between the two potentials are summarised as follows:
\begin{itemize}
    \item The functional form used for aPIP is significantly more complex than for the empirical force field, with the number of degrees of freedom being 5694 and 9, respectively. 
    \item The empirical force field includes only terms describing the interactions between atoms joined together by covalent bonds, whereas aPIP also naturally allows terms with nonbonded atoms (as long as they are within the spatial cutoff), e.g. the four-body HHHH term. 
    \item The empirical force field includes a five-body term (the angle-angle coupling) whilst the aPIP presented here is limited to four-body terms. 
\end{itemize}

Table \ref{table:rmse_test_train} shows that the energy RMSE of the empirical force field for methane decreases significantly as the body order is increased up to 3B; 4B and 5B terms each bring less than 15\% improvement. Compare this with the case of aPIPs (Fig.~\ref{fig:methane_no_basis_functions}), where a decrease in error of almost a factor of 50 is possible when going from 3B to 4B. 
The final test set errors are about 40 times larger for the empirical force field than for aPIP, at both 300K and 1500K. This strongly suggests that it is the constraints of the functional form, {\em rather than low body order} that limits the accuracy of empirical force fields. Note however, that the training and test set errors of the empirical force field are nearly the same, even though no regularisation was used in its fit---the need for careful regularisation is the price one pays for introducing enormous flexibility in the functional form.

\subsubsection{Comparison to high dimensional methods}

The aPIPs basis was introduced as a way to bridge the gap between empirical force fields and recent high dimensional machine learning based approaches. Therefore it is important to ask whether the limited body order aPIPs basis can reach the high accuracy of the ML methods, and at what computational cost. While we leave the detailed comparative study (including training and testing all methods on exactly the same data sets, optimising hyperparameters and computational efficiency of aPIPS, etc.) to future work, we can broadly answer in the affirmative. Methane and N-methyl acetamide (NMA) have both been fitted with the PIPs of Braams and Bowman. In Ref.~\cite{Nandi2019-bl}, energy RMSE of 0.4 meV was achieved on the training set of 1000 methane structures.
For NMA, Ref.~\cite{Qu2019-vb} gives the errors as 3.32~meV %
for the energy with the PIPs %
while the RMSE of the ``fragment method 2'' in that paper was 4.25~meV. %

For the benzene and ethanol molecules, the performance of aPIP can be compared to the sGDML~\cite{Chmiela2018}. The energy RMSEs achieved there for a 500K test set are 5.2~meV and 3.9~meV, for benzene and ethanol, respectively. The corresponding errors for the 4-body aPIPs for the 1500K test set are about 30\% higher.

\subsubsection{Bond, Angle and Dihedral Energy Scans}
We now demonstrate that the combination of low body order and regularisation results in smooth potential energy surfaces up to very high energies, several eV higher than the equilibrium energy, which are never encountered in the 1500K training and test sets. We performed bond, angle and dihedral scans for each molecule. The full set of results are given in Supplementary Information S1. %

\begin{figure}[ht]
    \centering
    \includegraphics[width=3.2in]{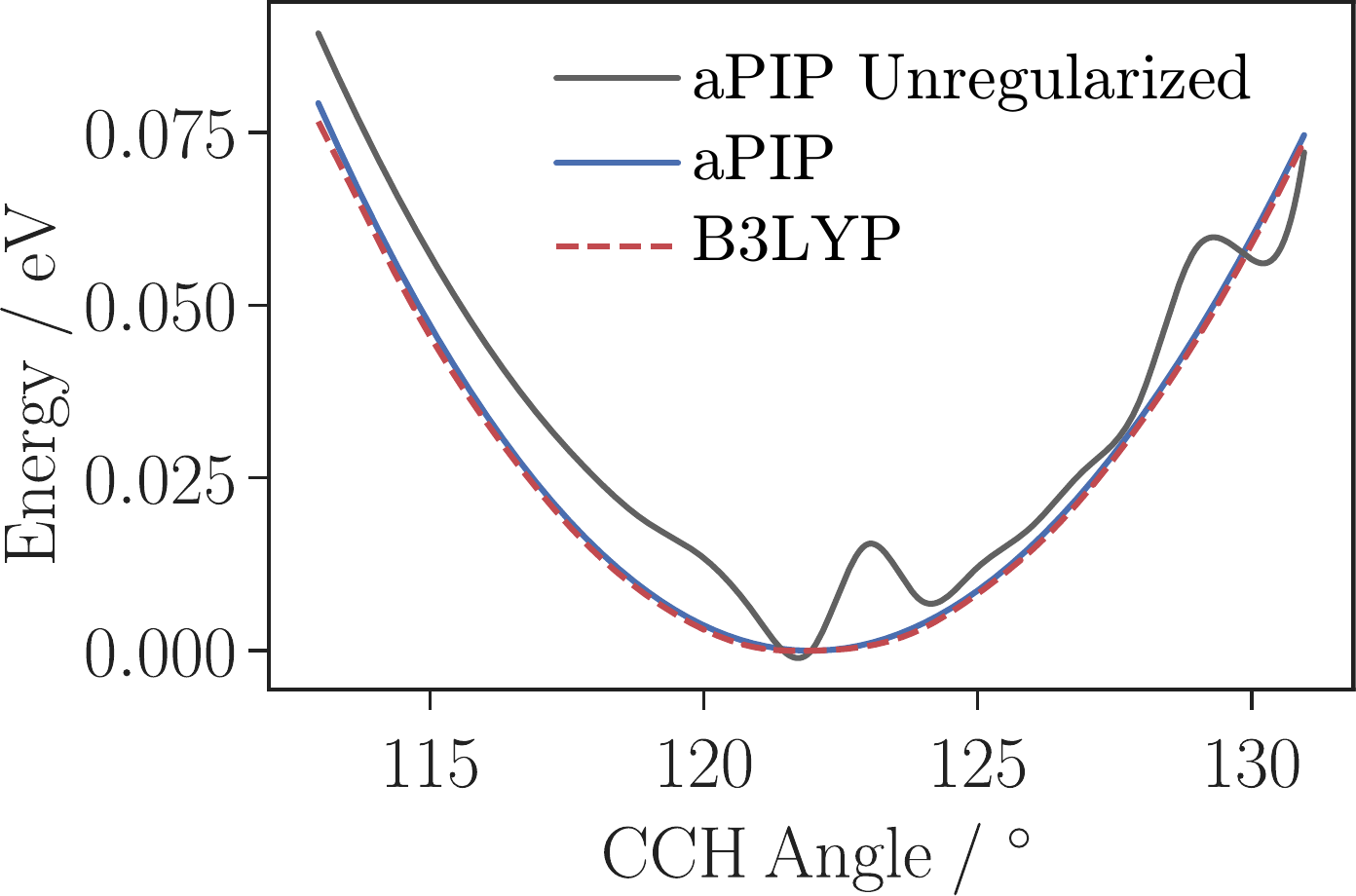}
    \caption{\label{fig:Butene_no_reg}Energy curve for a C--C--H angle in the molecule butene. Results for aPIP with and without regularization are shown. %
    } 
    
\end{figure}

The bond and angle aPIP energy scans show excellent agreement with the DFT results. The greatest differences between the aPIP and DFT scans occur for the hexane, butadiene and butene, which have more varied interactions and bonding present. However, even for these three molecules the energy scans are very well recreated. This level of accuracy for aPIP is in large part due to regularization. Figure \ref{fig:Butene_no_reg} demonstrates this with a C--C--H angle energy scan for butene. The regularised aPIP curve with regularization exactly follows the DFT energy scan, whilst the unregularised aPIP fit results in unphysical oscillations. 

We also calculated the energy curve for dilating adamantane~\cite{Duin2001}. Instead of just calculating the energy with the change in length of one individual bond, this test involves the uniform expansion of all the C--C bonds in adamantane. Again, a very close match with the DFT result is achieved. 
Note that adamantane has 26 atoms, this is over twice the size of N-methylacetamide, the largest molecule fit with PIPs~\cite{Qu2019-vb}. The fragmentation approach developed in Ref.~\citenum{Qu2019-vb} would not be suitable for adamantane, as the cyclic structure means that unambigous fragments do not exist and therefore, without the reduction in body order achieved with aPIPs, it would not be possible to create a PIP potential for this molecule. 

\begin{figure}[ht]
    \centering
    \includegraphics[width=3.2in]{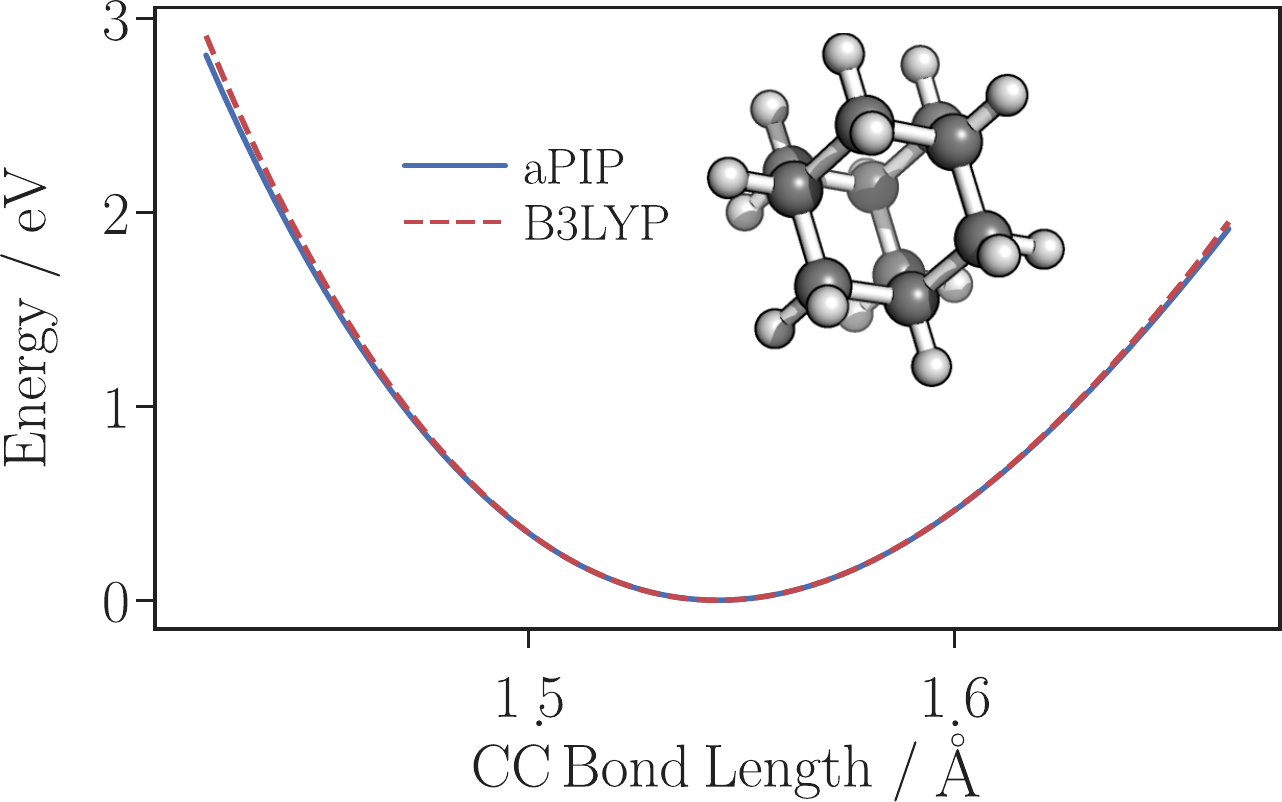}
    \caption{\label{fig:adamantane_all}The energy curve for dilating adamantane, this is the uniform increase of the C--C bond length.%
    }
    
\end{figure}

\begin{figure*}[ht]
    \centering
    \includegraphics[width=6.98in]{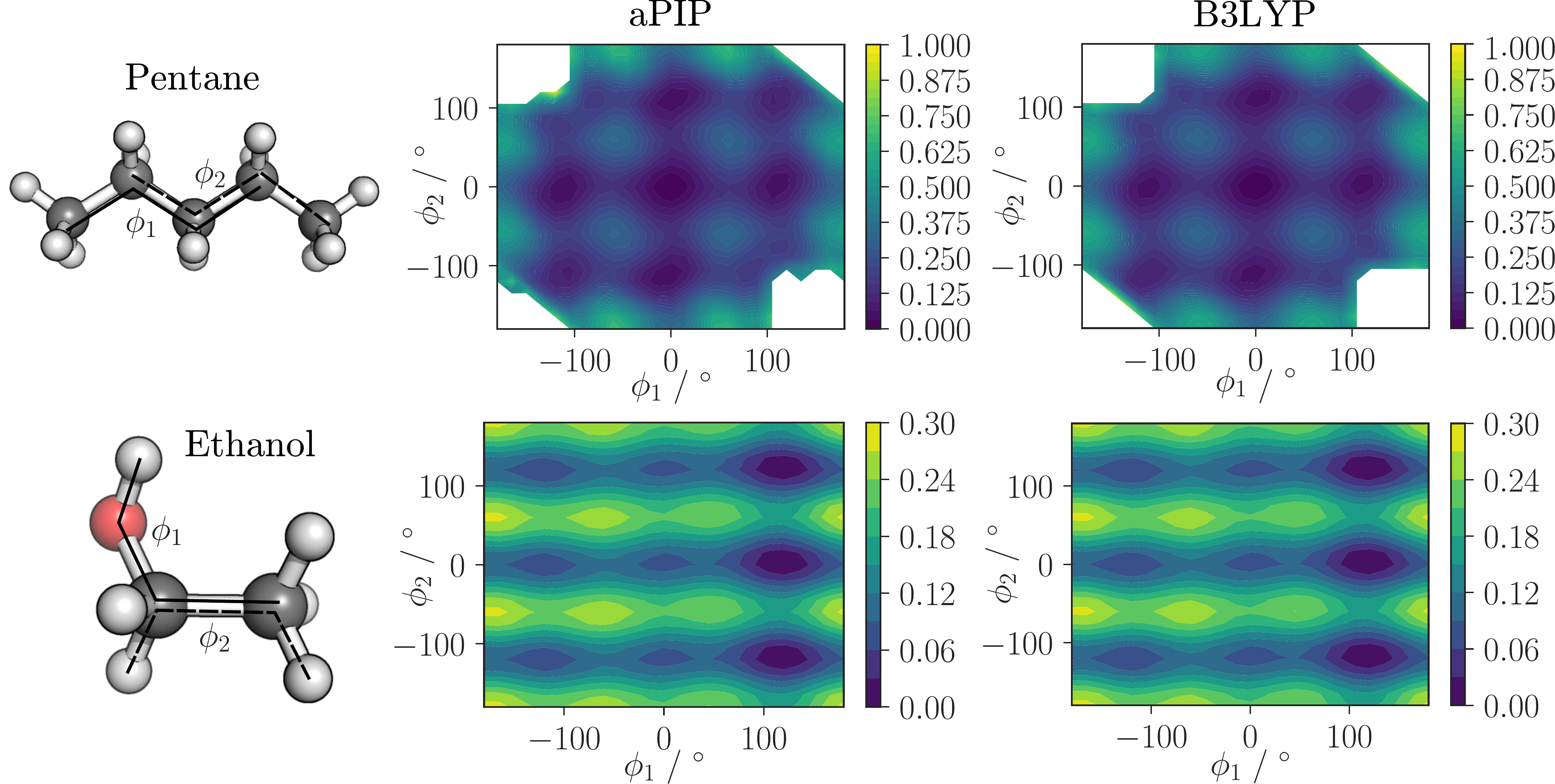}
    \caption{\label{fig:torsional_2d}The 2D dihedral angle energy scans for pentane and ethanol. The dihedral angles changed are marked on the molecule. The energy units are eV and values above 1.0~eV are not shown on the plot. %
    }
    
\end{figure*}

Accurate dihedral energy scans are an essential criterion for any organic molecular force field, and so a variety of dihedral angle energy scans for the aPIP potentials are shown in S1. The DFT energy is generally well reproduced. Hexane again shows slight discrepancy, with a RMSE of 26.7~meV, similar to the RMSE on the 300K test set. 

\begin{figure}[ht]
    \centering
    \includegraphics[width=3.20in]{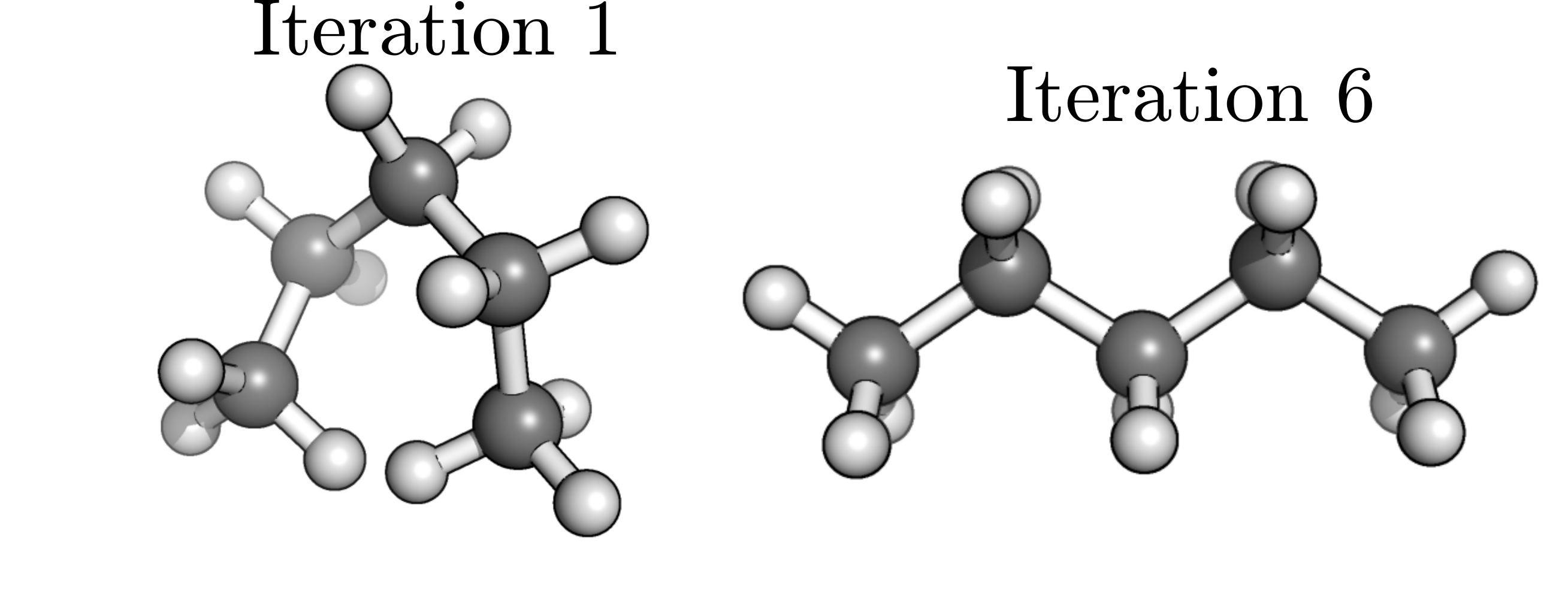}
    \caption{\label{fig:min_CCCCC}The lowest energy structure in the pentane 2D dihedral energy scan with the aPIP pentane potential at iteration 1 and at iteration~6. %
    }
    
\end{figure}

Further tests of aPIP's ability to reproduce dihedral effects is shown in the two dimensional maps of Fig.~\ref{fig:torsional_2d}. For both pentane and ethanol, minimum and maximum energy barriers are reproduced and the high energy regions in the DFT pentane map also occur in the aPIP map.
One of the interesting points to note about the aPIP pentane potential is that before the Sobol structures are added to the training set through the iterative fitting process,  the high energy regions of the 2D energy scan contain an unrealistically low energy structure, as shown in Fig.~\ref{fig:min_CCCCC}. 
At the first iteration of the fitting algorithm the RMSE for the 2D scan is 159~meV, whilst at the last iteration the RMSE is 36.6~meV, with the remaining error primarily due to the structures in the high energy region.

\begin{figure}[ht]
    \centering
    \includegraphics[width=3.20in]{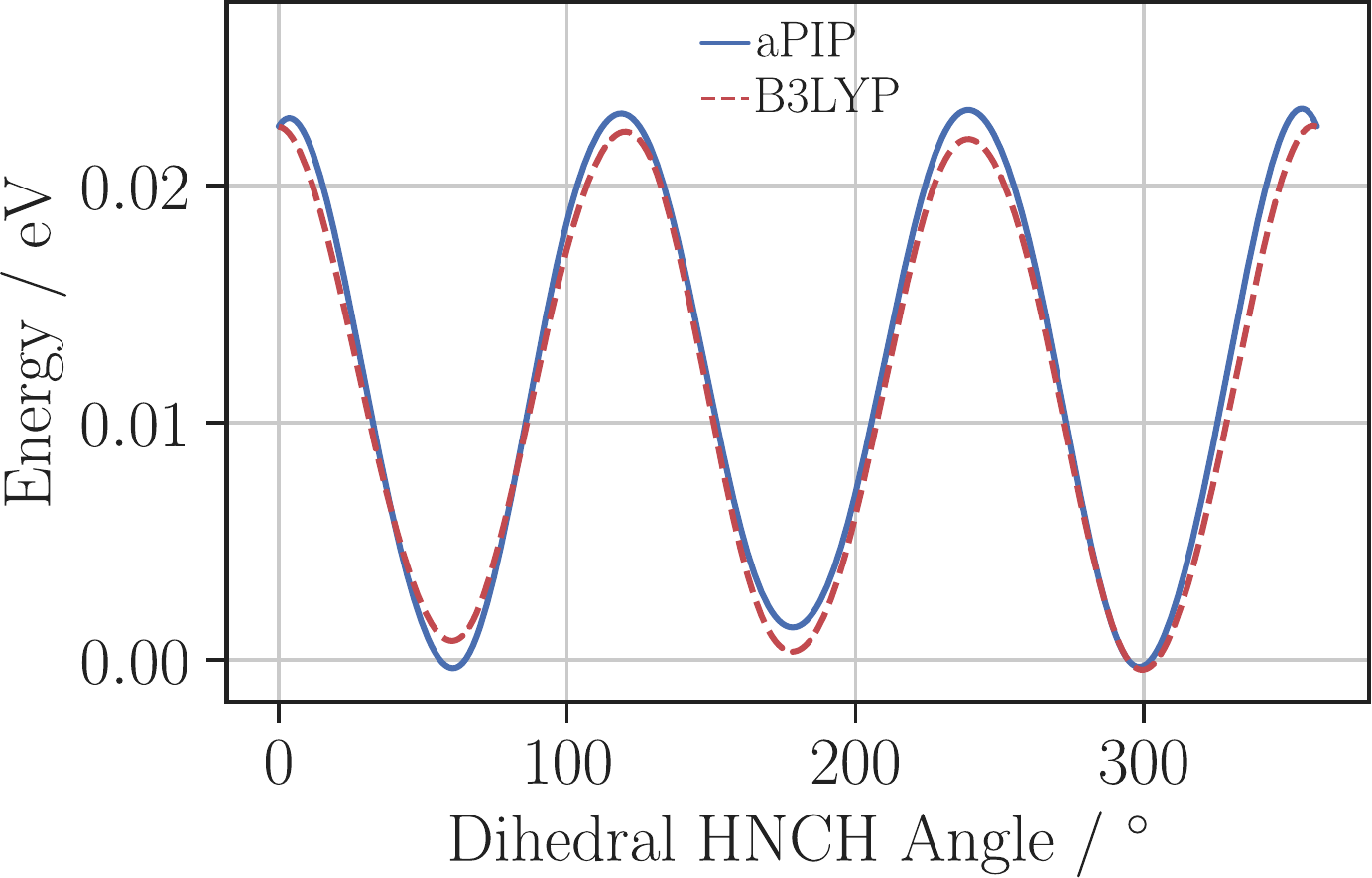}
    \caption{\label{fig:dihedral_NMA}The dihedral energy scan for the methyl group attached to the N--C bond of NMA. %
    }
    
\end{figure}

As a final example of the dihedral energy scan recreation, we show the energy scan of the methyl group attached to the N--C bond in NMA in Figure \ref{fig:dihedral_NMA}. The DFT energies and barrier heights are very well reproduced, and this along with the graphs shown in Supplementary Information S1 demonstrate the success of aPIPs for a molecule with four different element types.  

\subsubsection{Normal Mode Recreation }

Vibrational frequencies of molecules are regularly used as a measure of the accuracy of a force field. Empirical force fields with Class I functional forms, which have harmonic bond/angle terms and no coupling terms i.e. AMBER or OPLS, can achieve an error in the recreation of frequencies of approximately 50 ${\rm cm}^{-1}$~\cite{Allen2018} (mean absolute error, MAE) whilst Class II force fields, which include anharmonic and coupling terms in their functional forms, can achieve an MAE around 24~cm$^{-1}$~\cite{Ewig2001}. 

\begin{figure}[ht]
    \centering
    \includegraphics[width=3.20in]{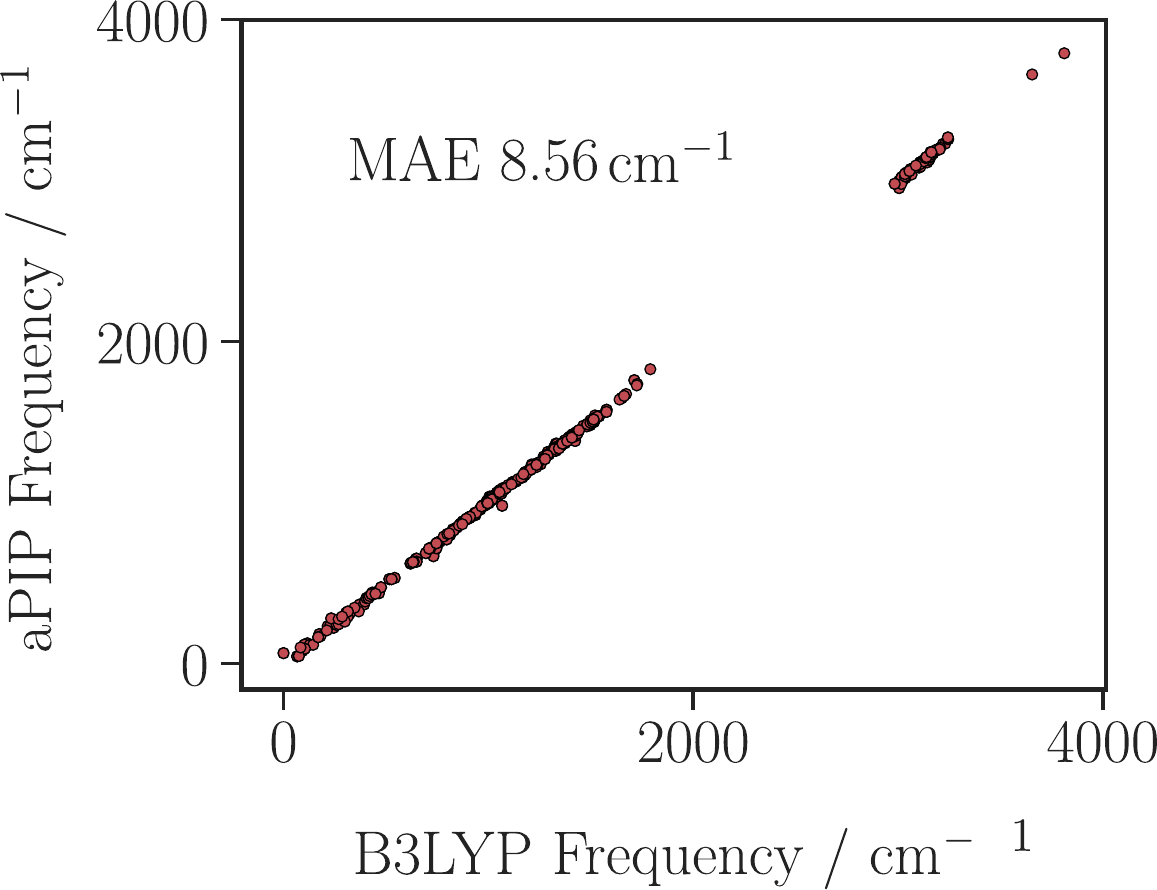}
    \caption{\label{fig:freq_all}The DFT and aPIP normal mode frequencies for the set of 14 molecules. The mean absolute error (MAE) for the normal mode recreation is 8.56 cm$^{-1}$.
    }
    
\end{figure}

The normal mode recreation for each individual molecule is given in S1 with all the DFT and aPIP frequencies for the molecules tested shown in Fig.~\ref{fig:freq_all}. With a MAE of 8.56 cm$^{-1}$ for the full set of molecules, aPIP recreates the normal mode frequencies with an accuracy that is far superior to empirical force fields. The individual molecules figures in S1 show that ethene has the highest MAE (16.6 cm$^{-1}$) whilst methane has a very low error with a MAE of just 0.647cm$^{-1}$.

\subsection{Combined Molecule Potentials}
\label{sec:combined_fits}

In this section, we show that the aPIP framework allows multiple molecules to be fit simultaneously, just as empirical force fields do.  This is in contrast to some high dimensional methods such as PIPs and sGDML.

\begin{table}
\begin{tabular}{c|c|c|}
\cline{2-3}
\multicolumn{1}{r|}{} & \multicolumn{2}{c|}{\textbf{Frequency }} \\
 & \multicolumn{2}{c|}{\textbf{MAE (cm$^{-1}$)}} \\ \cline{2-3} 
\multicolumn{1}{l|}{\textit{}} & \multicolumn{1}{l|}{\textit{Individual}} & \multicolumn{1}{l|}{\textit{Combined}} \\ \hline
\multicolumn{1}{|c|}{\textbf{Methane}} & 0.647 & 10.76 \\
\multicolumn{1}{|c|}{\textbf{Ethane}} & 4.93 & 10.76 \\
\multicolumn{1}{|c|}{\textbf{Propane}} & 4.8 & 13.71 \\
\multicolumn{1}{|c|}{\textbf{Butane}} & 10.65 & 16.88 \\
\multicolumn{1}{|c|}{\textbf{Pentane}} & 11.04 & 15.77 \\
\multicolumn{1}{|c|}{\textbf{Hexane}} & 15.06 & 17.72 \\
\multicolumn{1}{|c|}{\textbf{Heptane}} & - & 19.69 \\
\multicolumn{1}{|c|}{\textbf{Octane}} & - & 21.67 \\ \hline
\end{tabular}
\caption{\label{table:freq_combined_indiv}The normal mode recreation errors for the combined and individual aPIP potentials for a set of short linear alkanes. The training set includes the linear alkanes up to hexane. }

\end{table}

Table \ref{table:rmse_test_train} shows the testing and training RMSE for the aPIP model fitted to the combined training set of linear alkanes up to hexane. This can be compared to results for the individual aPIP fits in the same table. For alkanes with up to four carbon atoms, the individual molecule fits are superior. However, even for these four molecules the combined fit RMSE remains below 3~meV/atom on the high temperature test set. For pentane and hexane the combined fit gives the same or better levels of accuracy compared with the individual fits. Additionally, closer agreement between the training and test set RMSE is observed for the combined fit. This is due to the increase in the number and diversity of structures in the training set which further reduces overfitting.

Table \ref{table:freq_combined_indiv} showing the normal mode recreation exhibits a similar trend to the RMSE results. The error for shorter linear alkanes is lower with the individual aPIP potentials, but as the alkanes become longer the difference between the individual and combined aPIP potential errors decreases. The error in the combined molecule fit is still far below the typical errors expected from empirical force fields. 

\begin{figure}[h!]
    \centering
    \includegraphics[width=3.20in]{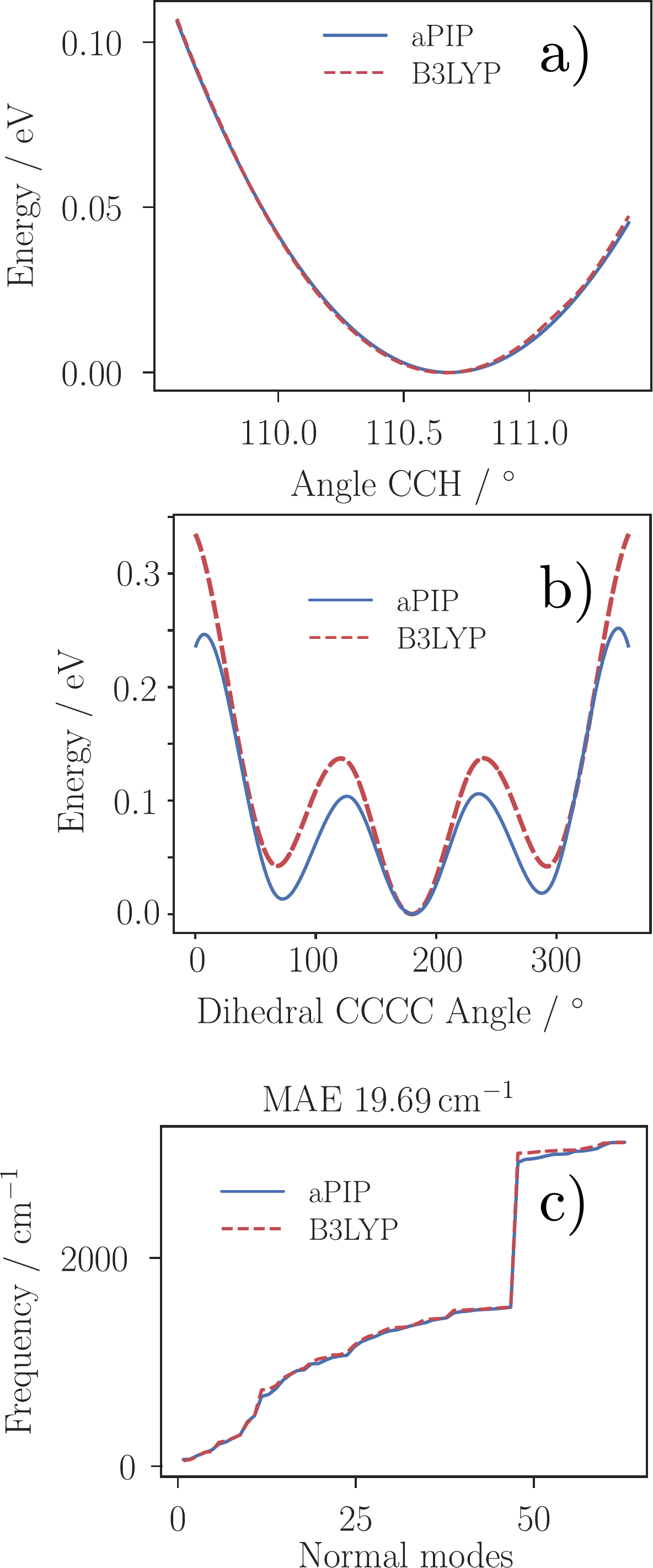}
    \caption{\label{fig:hept_ex} a) CCH angle and b) CCCC dihedral energy scans and normal mode recreation for heptane, which was not the training set. }
    
\end{figure}

The extrapolation capabilities of the combined potential to molecules not included in the fitting set are also demonstrated in Tables~\ref{table:rmse_test_train}, \ref{table:freq_combined_indiv} and Fig.~\ref{fig:hept_ex}. 
The RMSE for the 300~K testing set increases for the heptane and octane molecules, but stays below 2~meV/atom. 
The 1500~K test set RMSE shows a greater increase and demonstrates the need for larger data sets and possibly larger cutoffs. The normal mode recreation error for heptane and octane (Table \ref{table:freq_combined_indiv}) remains acceptable, with the error increasing only by 3.96~cm$^{-1}$ from hexane to octane. Figure \ref{fig:hept_ex} examines the energy scans for heptane. The CCH energy scan is  reproduced very well and the overall shape of the CCCC dihedral energy scan is also reasonable. However, the trans-gauche dihedral energy barrier is 0.03eV lower than the corresponding DFT value.%

Currently, the best transferable high dimensional force field is ANI~\cite{Smith2017,Smith2018-less,Smith2019}. While a detailed comparative study is left for future work, the DFT version (ANI-1) gives RMSE errors on the GDB-11 database of about 2.9~meV/atom~\cite{Smith2017}, higher than what the combined 4-body aPIP fit achieves on our limited range of molecules.

\label{sec:combined_results}

\section{Discussion and Conclusion}

In this work, we have built on the ideas introduced in Ref.~\citenum{Van_der_Oord2019-td}, which reformulated the permutationally invariant polynomial basis for single element materials, and created potentials for organic molecules using the multi-element atomic permutationally invariant polynomial (aPIP) basis. We showcase potentials that restrict the body order (in the present case to four), and employ a bond-angle based coordinate system, cutoffs for large and small distances, a repulsive core, and regularize the least square fitting. These alterations allow  potentials for much larger molecules to be created and multiple molecules to be fit at once, in contrast to the original PIP framework. Additionally, by a combination of regularization and iterative training, the ``holes'' in the potential are eliminated, making them suitable for molecular dynamics (see the SI for examples).

The performance of the aPIP potentials, both individually fitted to organic molecules and simultaneously to a combined set, showed very good accuracy for a number of properties (e.g. a few meV  per atom error for the energy at 1500K), on a par with recent machine learning approaches. The speed of aPIP potentials is of course much slower than that of empirical force fields, but is the same order of magnitude as other ML potentials: typically on the order of 1 ms/atom. Fast implementations of polynomial bases exist (certainly for MTP~\cite{Shapeev2016-hn} and also ACE~\cite{Drautz:2019,Bachmayr2019-ec}) that will bring this time down further. 

Furthermore, the relatively small dimensionality of low body order terms coupled with well controlled regularisation results in smooth potentials and remarkable extrapolation properties. Returning to Fig.~\ref{fig:ch4dis}, we see that the aPIP dissociation curves of methane are smooth and qualitatively correct, even though the only data that informs the potential are near equilibrium geometries, and the isolated atom energies. The latter only ensures that the simultaneous removal of all four Hs gives the correct limit at infinite distance (black dashed line in the bottom panel), the rest is extrapolation. 

We have also outlined the relationships between the  approaches for making force fields. Although each have rather distinct assumptions and seemingly incompatible mathematical frameworks, it turns out that body ordered polynomials (either the aPIPs variety used in this work, or the atomic cluster expansion) form links between them. This points the way forward to creating potentials that do not require atom typing, can be reactive and  transferable, but remain highly accurate approximators of the Born-Oppenheimer potential energy surface. 

Building a comprehensive organic force field is a significantly larger undertaking, but our limited results already show that achieving high accuracy does {\em not} necessarily need nonlinear fitting such as neural networks or even kernel methods. This, which we consider the main point of this work, is at variance with what might be gleaned from recent trends in the literature. 

\begin{acknowledgements}
This work was partially supported by the French "Investissements d'Avenir" program, project ISITE-BFC (contract ANR-15-IDEX-0003). CO was supported by the Leverhulme Trust under Grant RPG-2017-191. 
\end{acknowledgements}

\bibliography{refs}




\section*{Supplementary material}

# Supporting information for:

# Atomic Permutationally Invariant Polynomials for

# Fitting Molecular Force Fields

Alice E. A. Allen,[*,†] Geneviève Dusson,[*,‡] Gábor Csányi,[*,†] and Christoph

Ortner[*,¶]

*Engineering Laboratory, University of Cambridge, Trumpington Street,Cambridge, CB2*

*1PZ, United Kingdom, Université Bourgogne Franche-Comté, Laboratoire de*

*Mathématiques de Besançon, UMR CNRS 6623, Besançon, France, and Mathematics*

*Department, University of British Columbia, 1984 Mathematics Rd, Vancouver, BC V6T*

*1Z2, Canada*

E-mail: aa840@cam.ac.uk; genevieve.dusson@math.cnrs.fr; gc121@cam.ac.uk;

c.ortner@warwick.ac.uk

---

[*]To whom correspondence should be addressed
[†]University of Cambridge
[‡]Université Bourgogne Franche-Comté
[¶]Warwick University

# Contents



# S1 Energy Scans

## S1.1 Alkanes

### S1.1.1 Adamantane

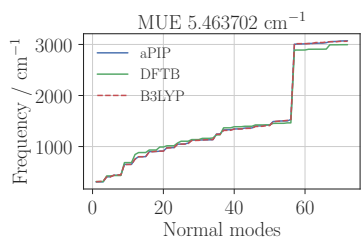

Figure S1: Normal Modes for Adamantane

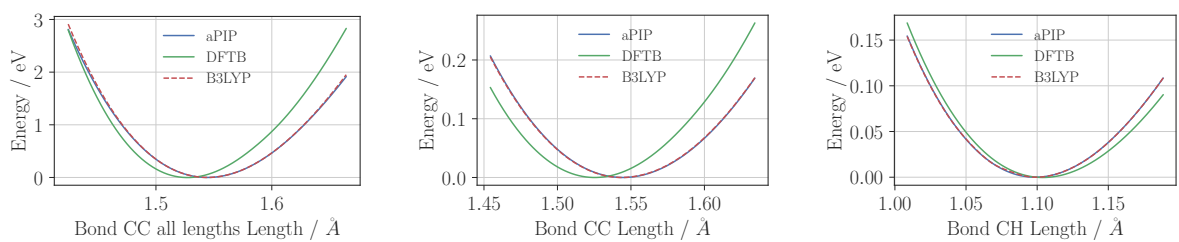

Figure S2: Bond Lengths for Adamantane

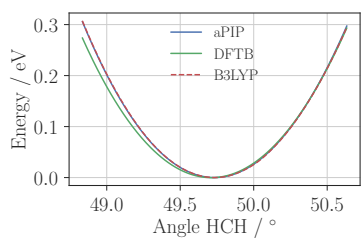

Figure S3: Angles for Adamantane

Training Set - 1000 structures, 1500K MD

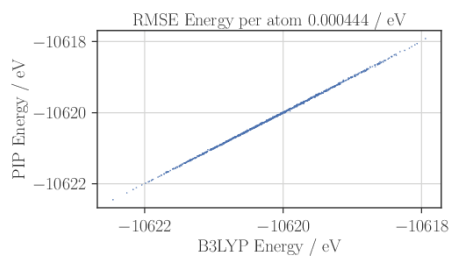

Test Set - 8000 structures, 1500K MD

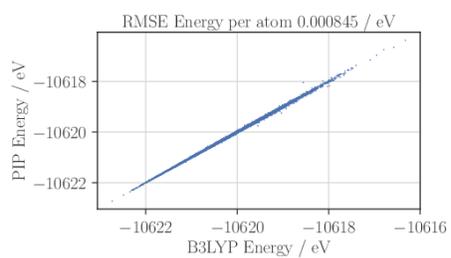

Test Set - 8000 structures, 300K MD

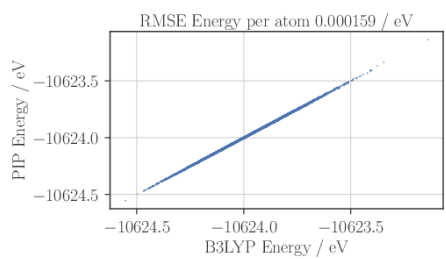

Figure S4: Comparison between the QM and PIP energies for Adamantane

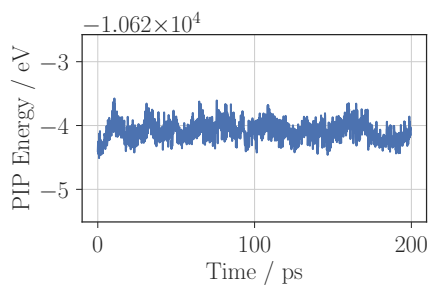

Figure S5: Energy over the course of a 300K MD simulation whilst using an aPIP potential for Adamantane

### S1.1.2 Butane

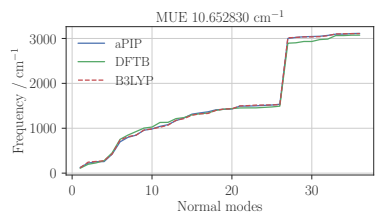

Figure S6: Normal Modes for Butane

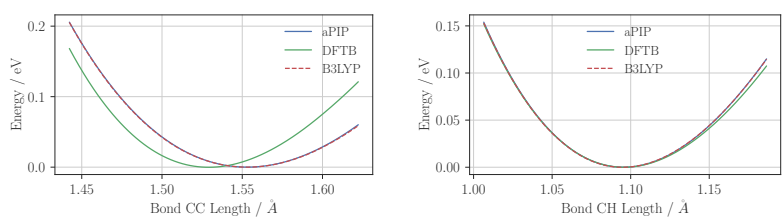

Figure S7: Bond Lengths for Butane

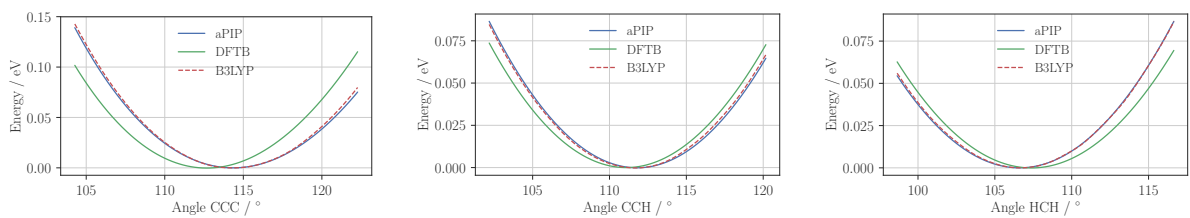

Figure S8: Angles for Butane

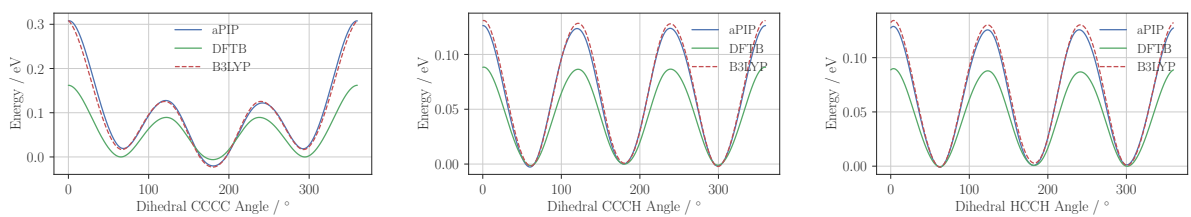

Figure S9: Dihedrals for Butane

Training Set - 1000 structures, 1500K MD

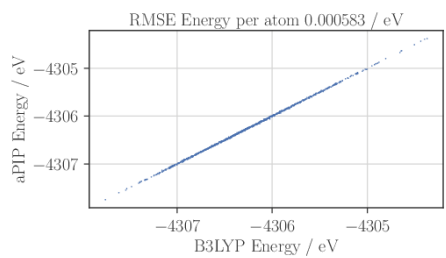

Test Set - 8000 structures, 1500K MD

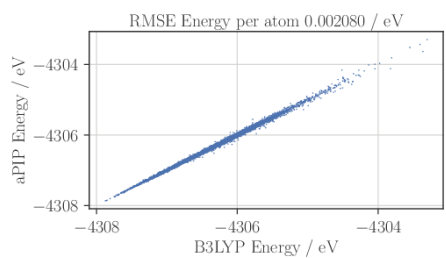

Test Set - 8000 structures, 300K MD

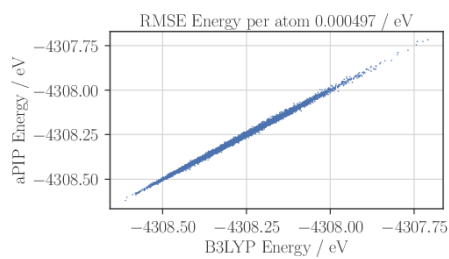

Figure S10: Comparison between the QM and PIP energies for Butane

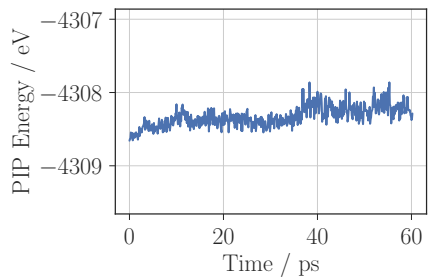

Figure S11: Energy over the course of a 300K MD simulation whilst using an aPIP potential for Butane

### S1.1.3 Ethane

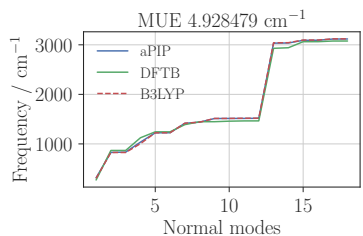

Figure S12: Normal Modes for Ethane

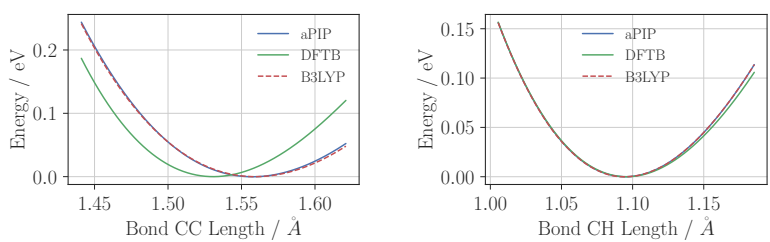

Figure S13: Bond Lengths for Ethane

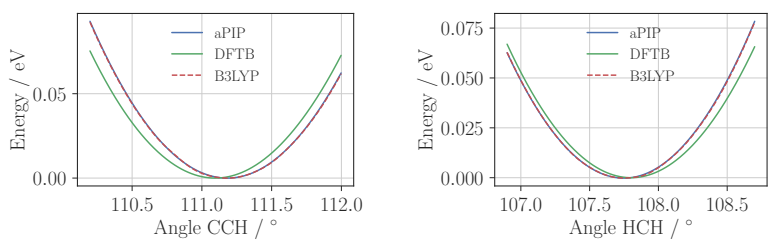

Figure S14: Angles for Ethane

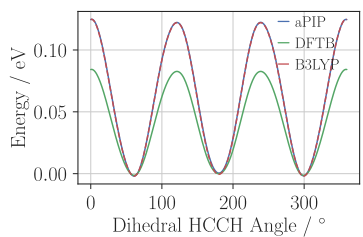

Figure S15: Dihedrals for Ethane

Training Set - 1000 structures, 1500K MD

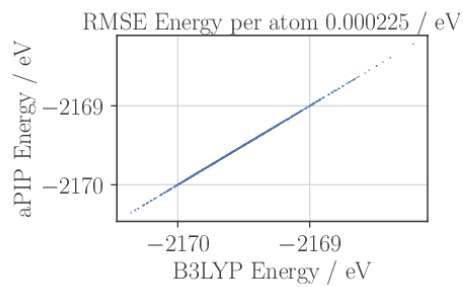

Test Set - 8000 structures, 1500K MD

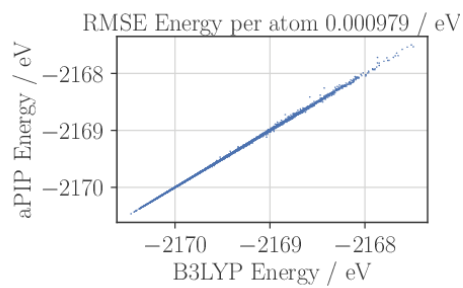

Test Set - 8000 structures, 300K MD

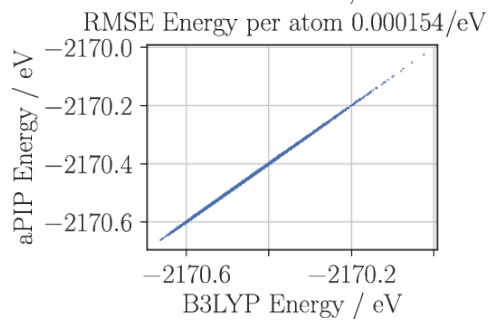

Figure S16: Comparison between the QM and PIP energies for Ethane

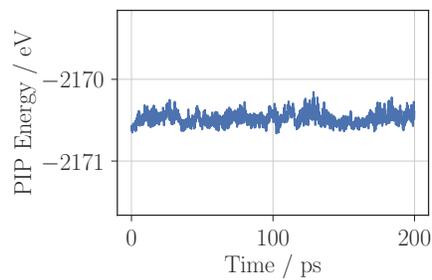

Figure S17: Energy over the course of a 300K MD simulation whilst using an aPIP potential for Ethane

## S1.1.4 Hexane

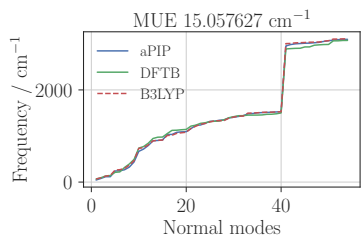

Figure S18: Normal Modes for Hexane

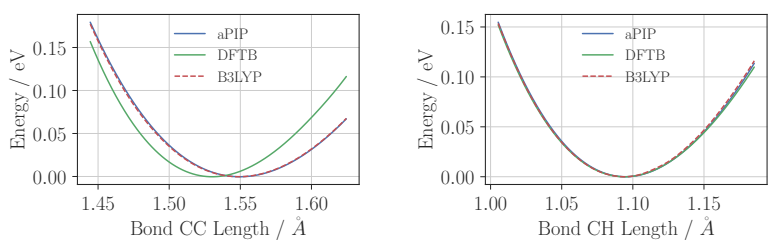

Figure S19: Bond Lengths for Hexane

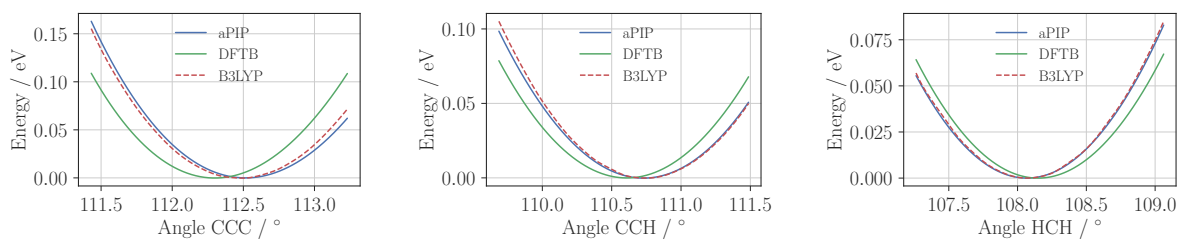

Figure S20: Angles for Hexane

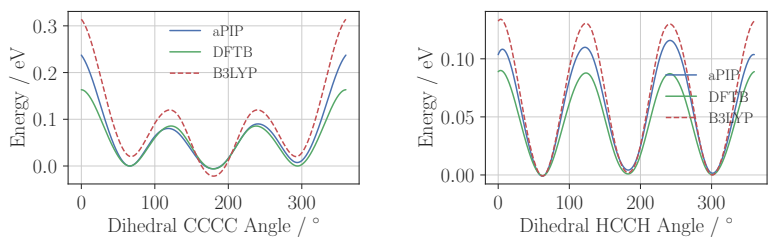

Figure S21: Dihedrals for Hexane

**Training Set - 1000 structures, 1500K MD**

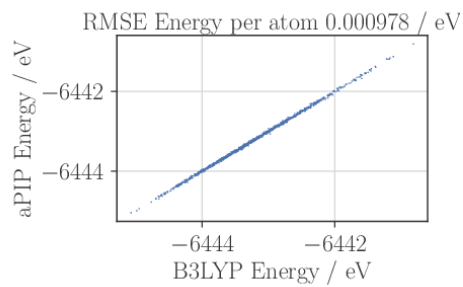

**Test Set - 8000 structures, 1500K MD**

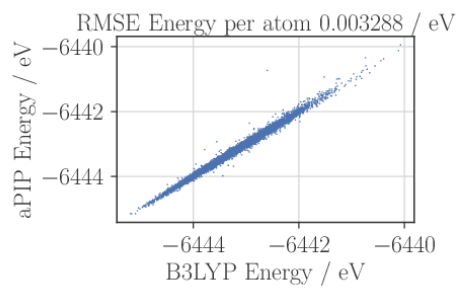

**Test Set - 8000 structures, 300K MD**

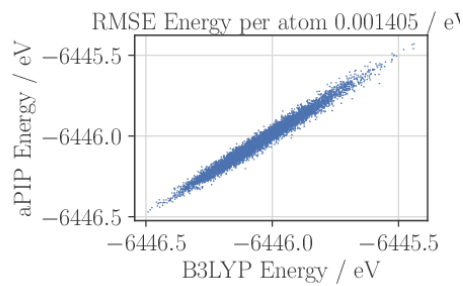

Figure S22: Comparison between the QM and PIP energies for Hexane

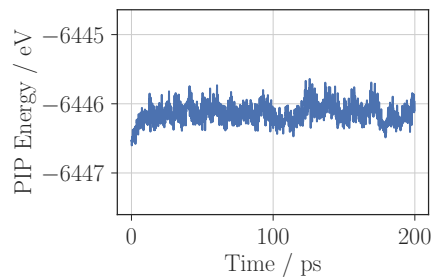

Figure S23: Energy over the course of a 300K MD simulation whilst using an aPIP potential for Hexane

### S1.1.5 Methane

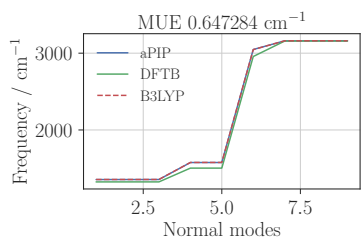

Figure S24: Normal Modes for Methane

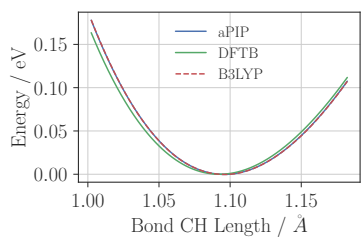

Figure S25: Bond Lengths for Methane

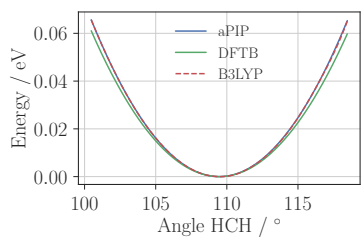

Figure S26: Angles for Methane

Training Set - 1000 structures, 1500K MD

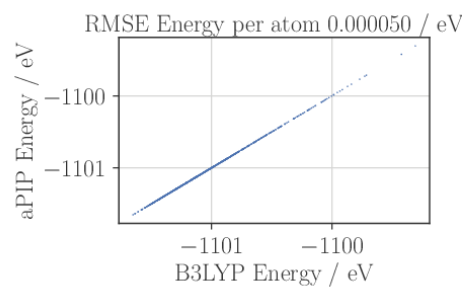

Test Set - 8000 structures, 1500K MD

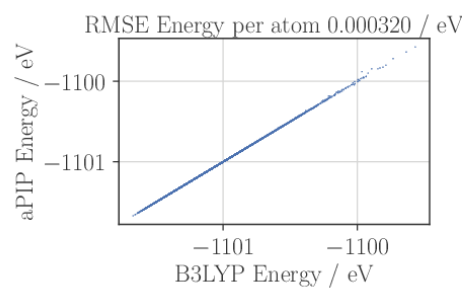

Test Set - 8000 structures, 300K MD

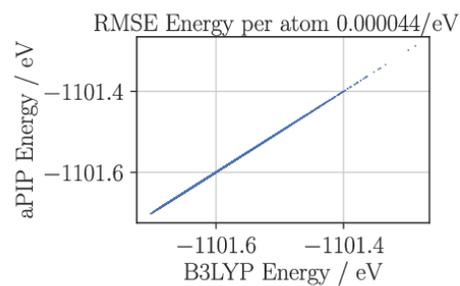

Figure S27: Comparison between the QM and PIP energies for Methane

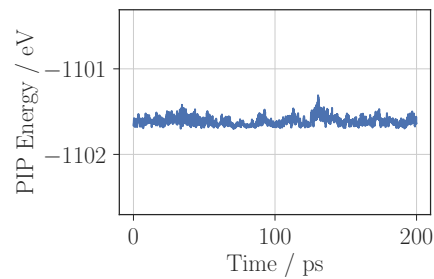

Figure S28: Energy over the course of a 300K MD simulation whilst using an aPIP potential for Methane

## S1.1.6   Pentane

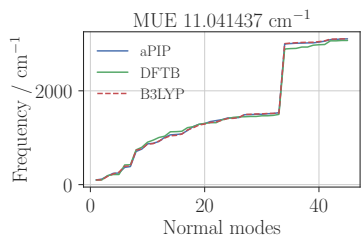

Figure S29: Normal Modes for Pentane

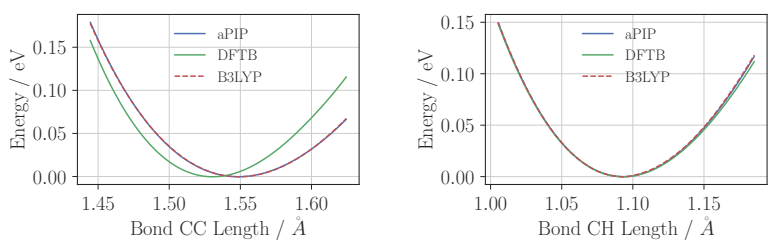

Figure S30: Bond Lengths for Pentane

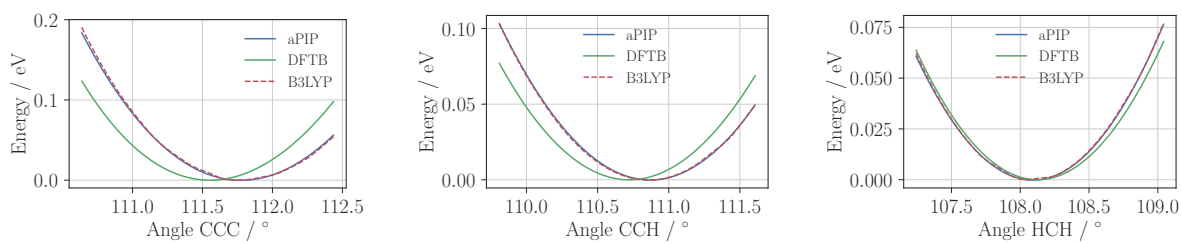

Figure S31: Angles for Pentane

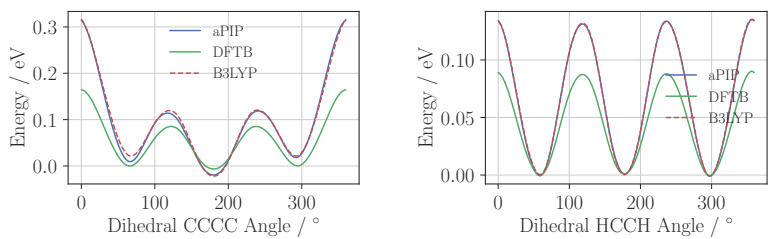

Figure S32: Dihedrals for Pentane

Training Set - 1000 structures, 1500K MD

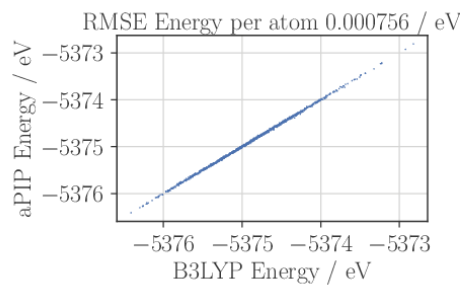

Test Set - 8000 structures, 1500K MD

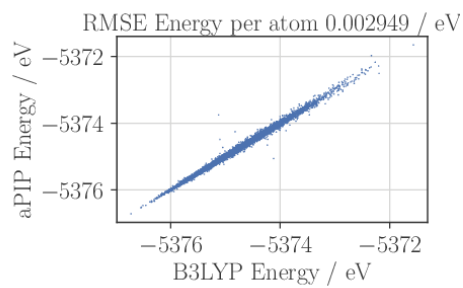

Test Set - 8000 structures, 300K MD

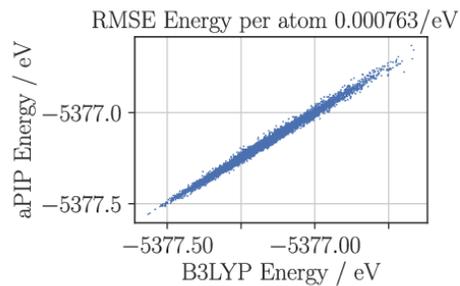

Figure S33: Comparison between the QM and PIP energies for Pentane

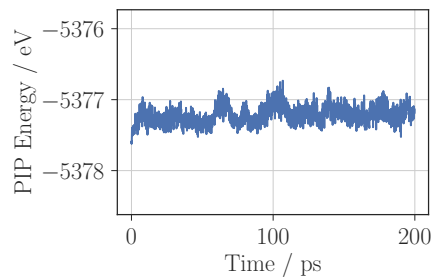

Figure S34: Energy over the course of a 300K MD simulation whilst using an aPIP potential for Pentane

### S1.1.7 Propane

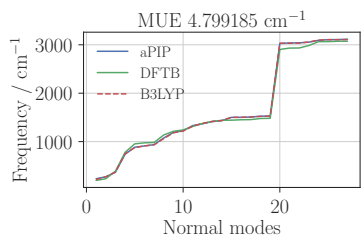

Figure S35: Normal Modes for Propane

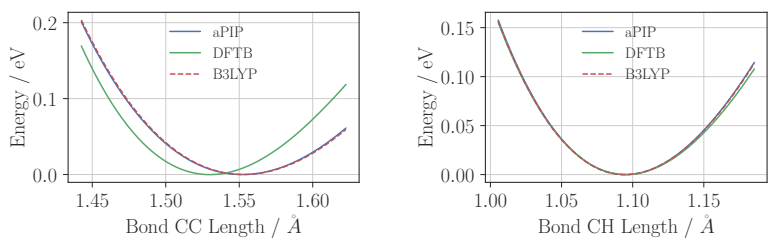

Figure S36: Bond Lengths for Propane

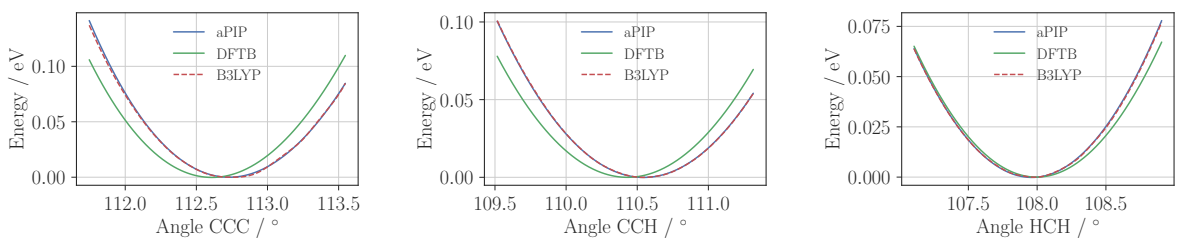

Figure S37: Angles for Propane

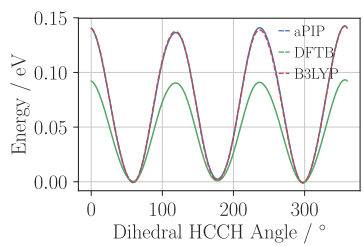

Figure S38: Dihedrals for Propane

Training Set - 1000 structures, 1500K MD

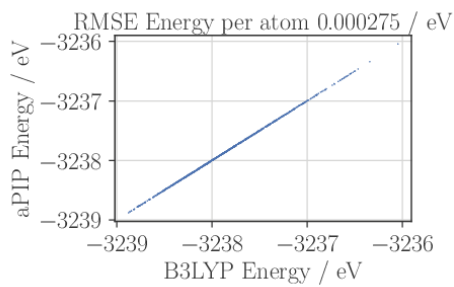

Test Set - 8000 structures, 1500K MD

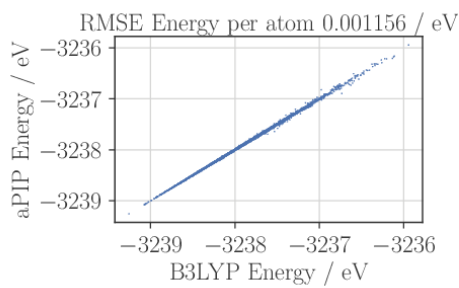

Test Set - 8000 structures, 300K MD

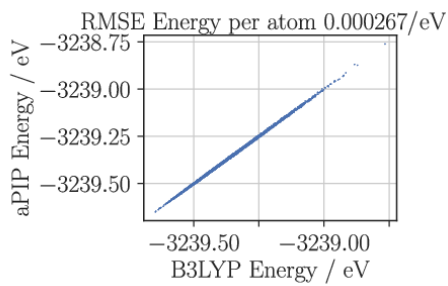

Figure S39: Comparison between the QM and PIP energies for Propane

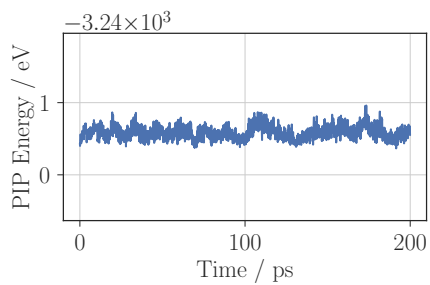

Figure S40: Energy over the course of a 300K MD simulation whilst using an aPIP potential for Propane

## S1.2 Alkenes

### S1.2.1 Butadiene

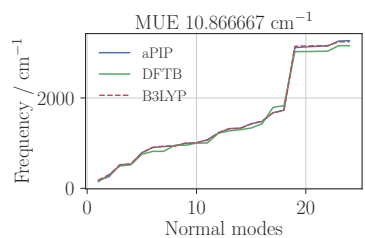

Figure S41: Normal Modes for Butadiene

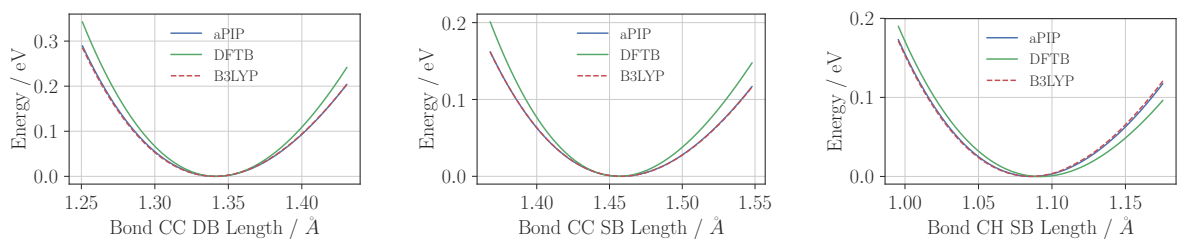

Figure S42: Bond Lengths for Butadiene

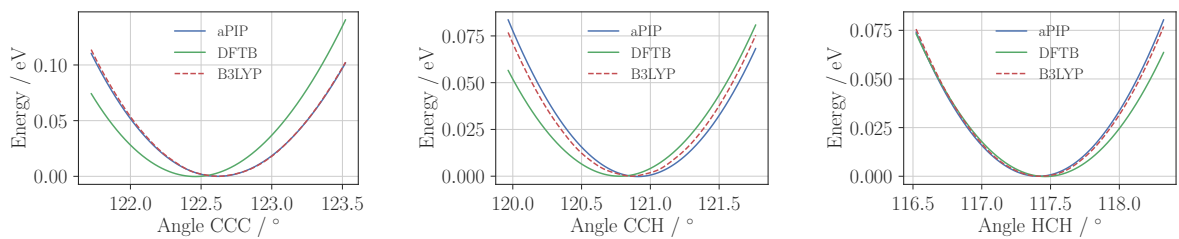

Figure S43: Angles for Butadiene

Training Set - 1000 structures, 1500K MD

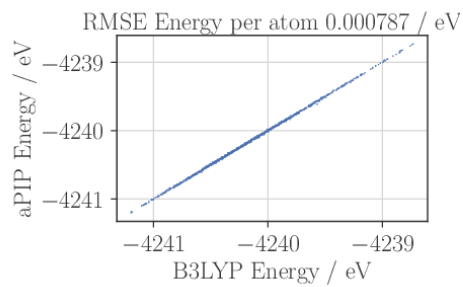

Test Set - 8000 structures, 1500K MD

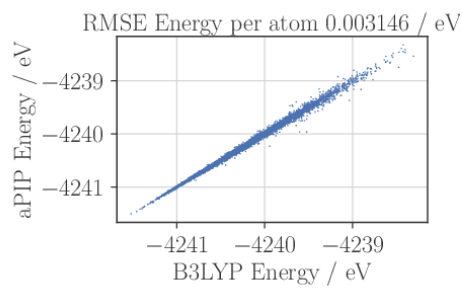

Test Set - 8000 structures, 300K MD

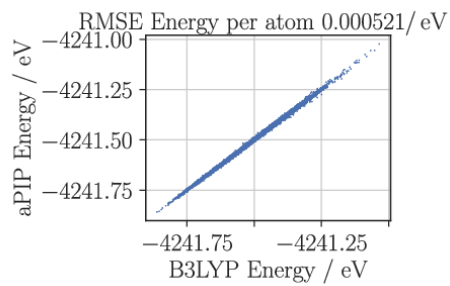

Figure S44: Comparison between the QM and PIP energies for Butadiene

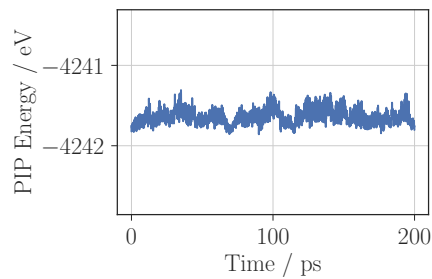

Figure S45: Energy over the course of a 300K MD simulation whilst using an aPIP potential for Butadiene

## S1.2.2 Butene

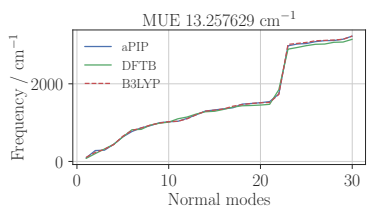

Figure S46: Normal Modes for Butene

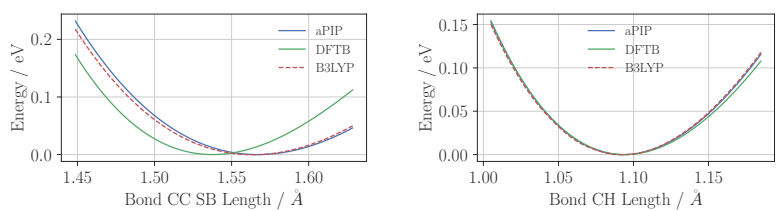

Figure S47: Bond Lengths for Butene

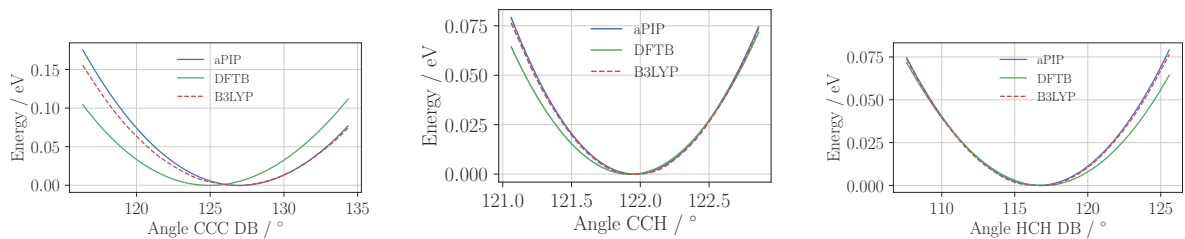

Figure S48: Angles for Butene

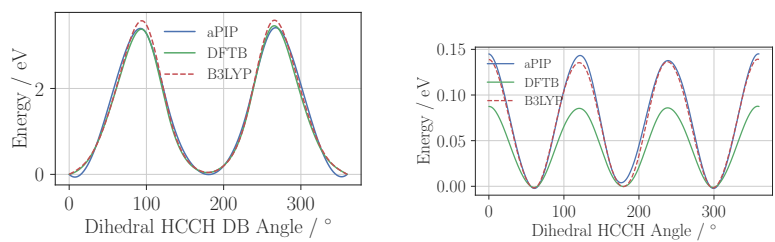

Figure S49: Dihedrals for Butene

Training Set - 1000 structures, 1500K MD

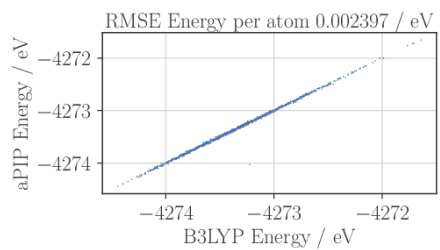

Test Set - 8000 structures, 1500K MD

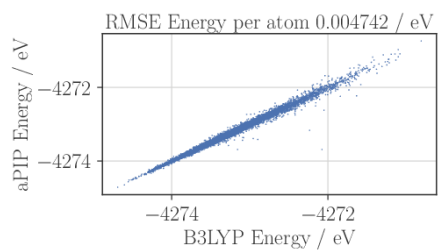

Test Set - 8000 structures, 300K MD

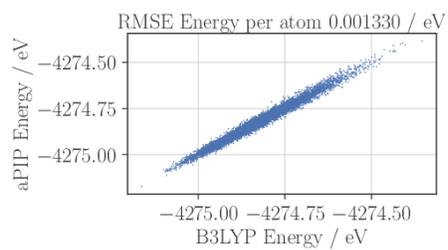

Figure S50: Comparison between the QM and PIP energies for Butene

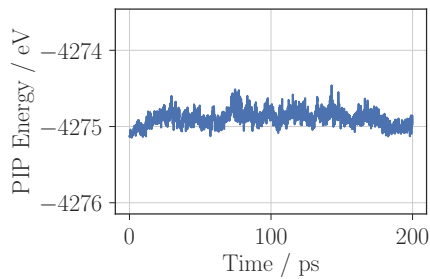

Figure S51: Energy over the course of a 300K MD simulation whilst using an aPIP potential for Butene

### S1.2.3 Ethene

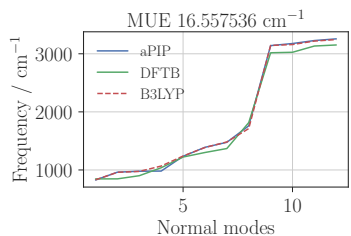

Figure S52: Normal Modes for Ethene

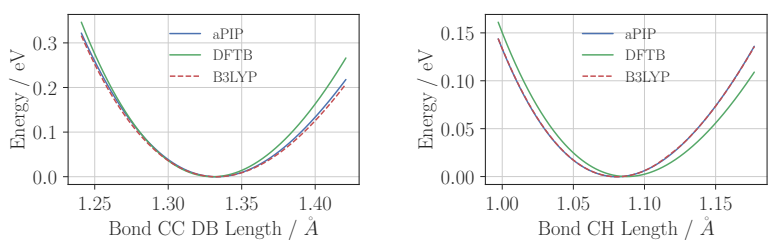

Figure S53: Bond Lengths for Ethene

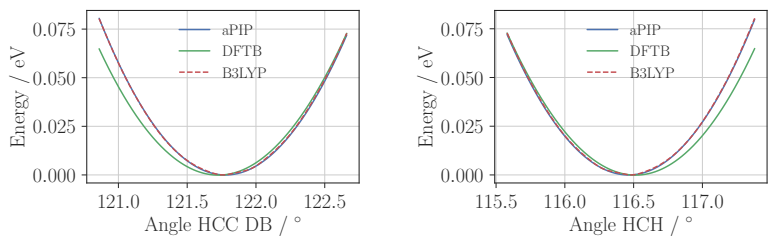

Figure S54: Angles for Ethene

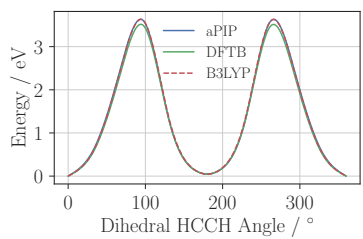

Figure S55: Dihedrals for Ethene

Training Set - 1000 structures, 1500K MD

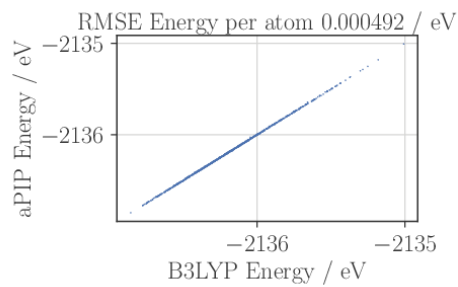

Test Set - 8000 structures, 1500K MD

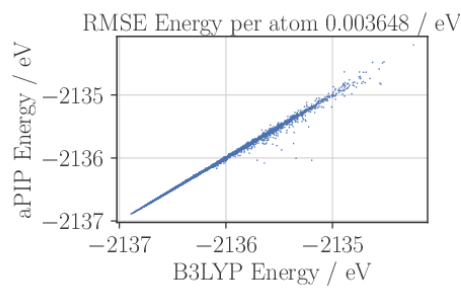

Test Set - 8000 structures, 300K MD

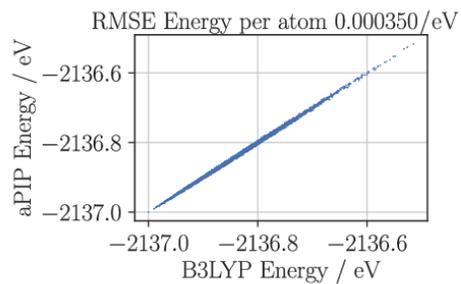

Figure S56: Comparison between the QM and PIP energies for Ethene

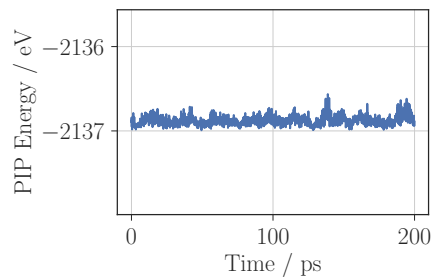

Figure S57: Energy over the course of a 300K MD simulation whilst using an aPIP potential for Ethene

## S1.3 Aromatic

### S1.3.1 Benzene

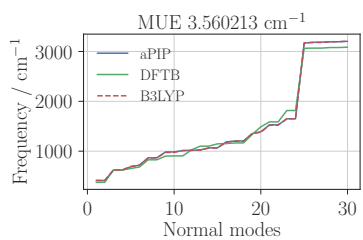

Figure S58: Normal Modes for Benzene

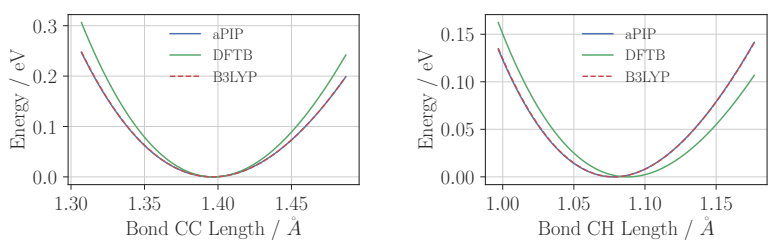

Figure S59: Bond Lengths for Benzene

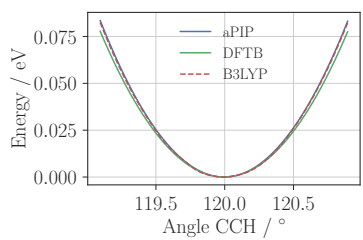

Figure S60: Angles for Benzene

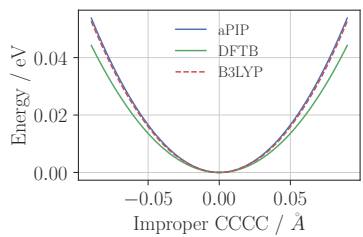

Figure S61: Impropers for Benzene

Training Set - 1000 structures, 1500K MD

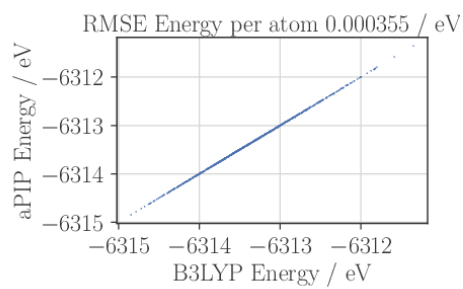

Test Set - 8000 structures, 1500K MD

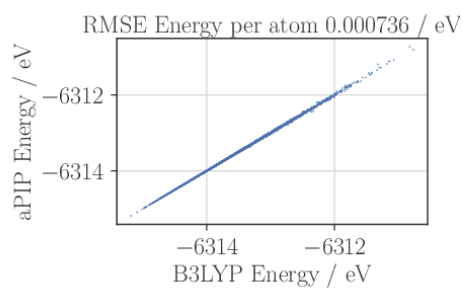

Test Set - 8000 structures, 300K MD

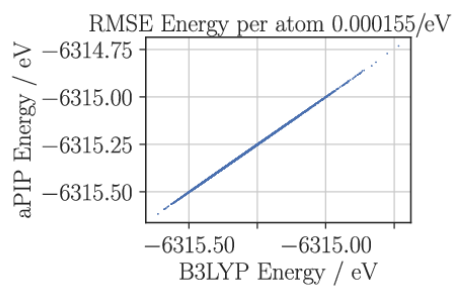

Figure S62: Comparison between the QM and PIP energies for Benzene

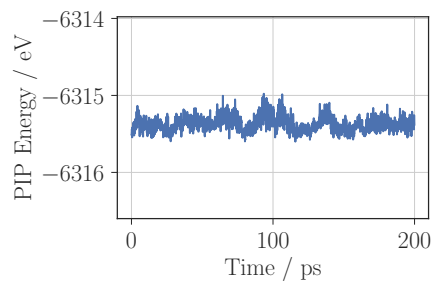

Figure S63: Energy over the course of a 300K MD simulation whilst using an aPIP potential for Benzene

## S1.3.2  Methylbenzene

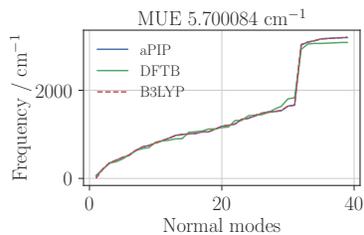

Figure S64: Normal Modes for Methylbenzene

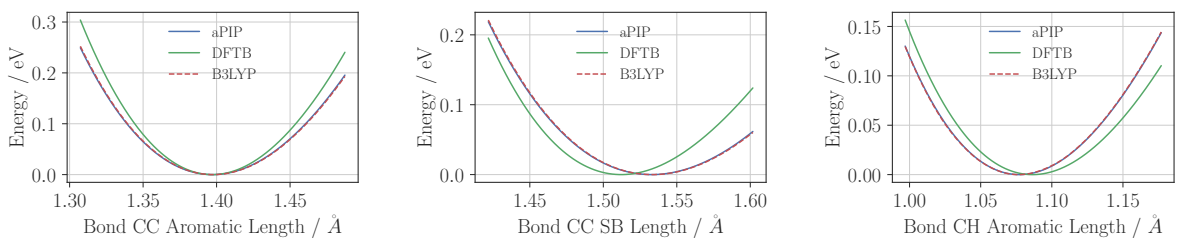

Figure S65: Bond Lengths for Methylbenzene

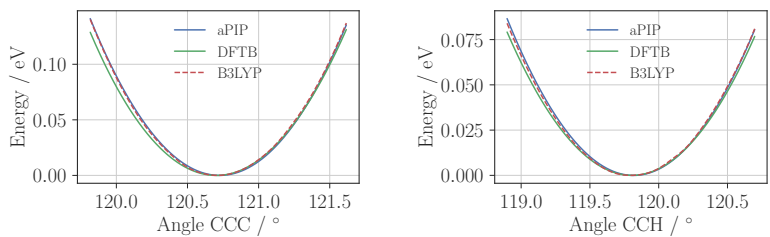

Figure S66: Angles for Methylbenzene

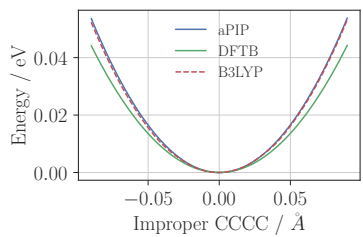

Figure S67: Impropers for Methylbenzene

Training Set - 1000 structures, 1500K MD

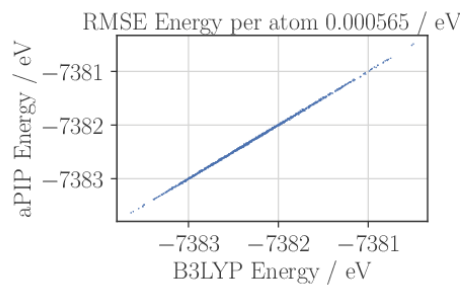

Test Set - 8000 structures, 1500K MD

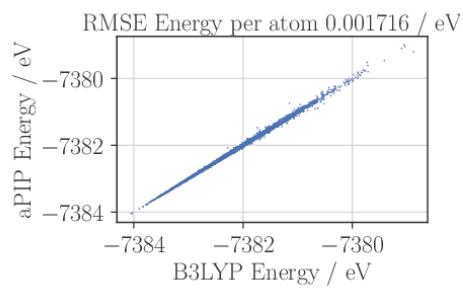

Test Set - 8000 structures, 300K MD

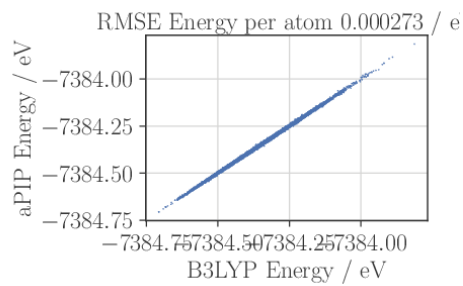

Figure S68: Comparison between the QM and PIP energies for Methylbenzene

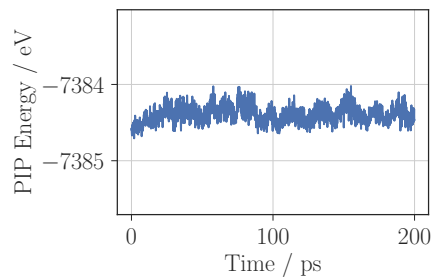

Figure S69: Energy over the course of a 300K MD simulation whilst using an aPIP potential for Methylbenzene

## S1.4 Other

### S1.4.1 Ethanol

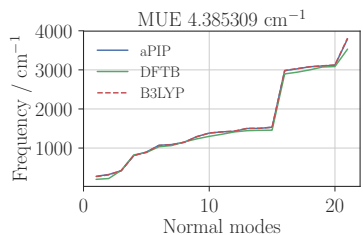

Figure S70: Normal Modes for Ethanol

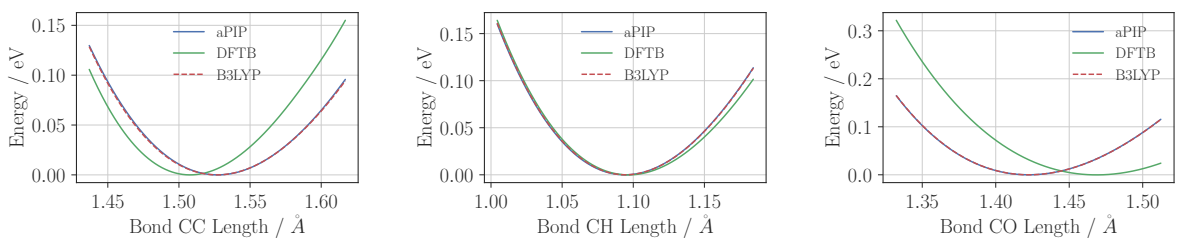

Figure S71: Bond Lengths for Ethanol

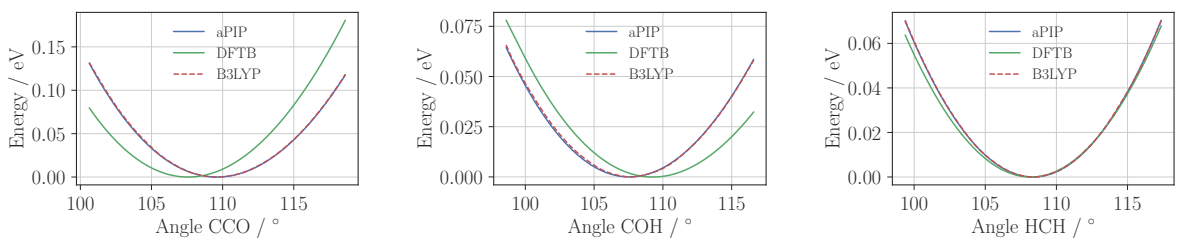

Figure S72: Angles for Ethanol

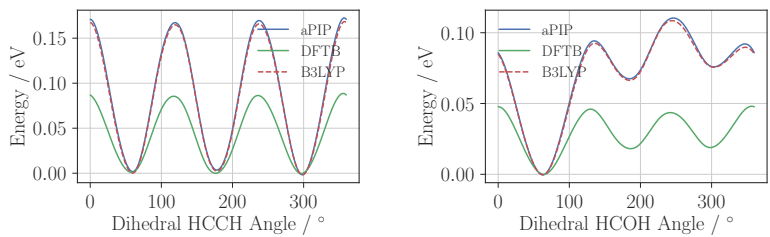

Figure S73: Dihedrals for Ethanol

Test Set - 8000 structures, 300K MD

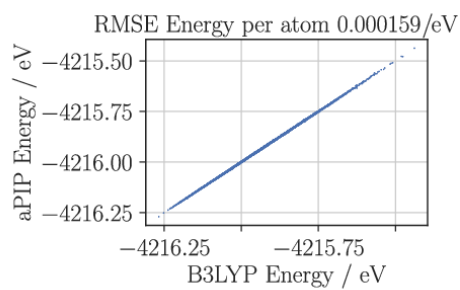

Training Set - 1000 structures, 800K MD

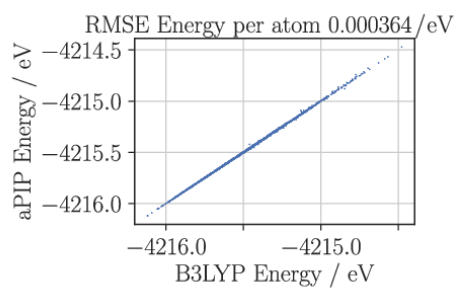

Test Set - 8000 structures, 800K MD

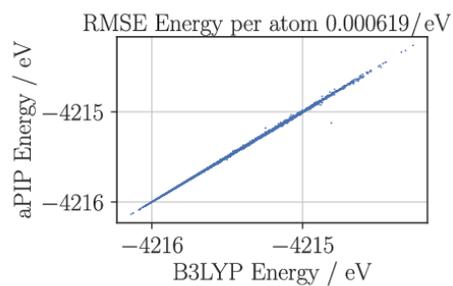

Figure S74: Comparison between the QM and PIP energies for Ethanol

## S1.4.2 NMA

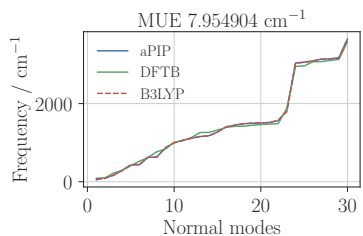

Figure S75: Normal Modes for NMA

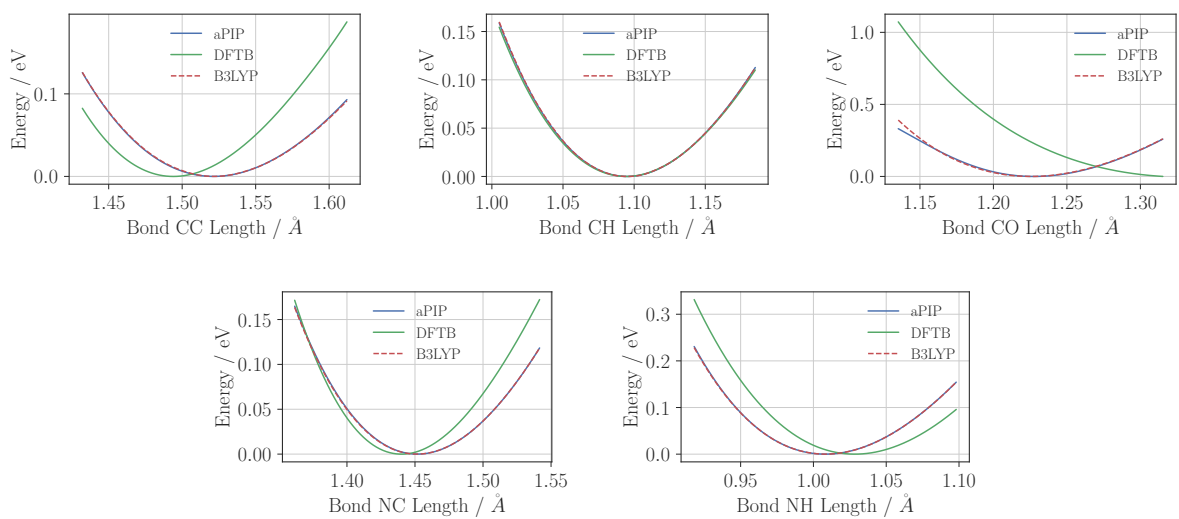

Figure S76: Bond Lengths for NMA

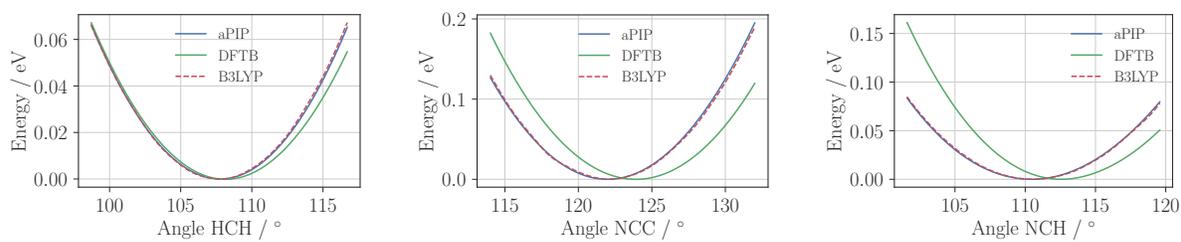

Figure S77: Angles for NMA

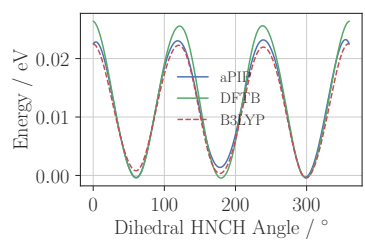

Figure S78: Dihedrals for NMA

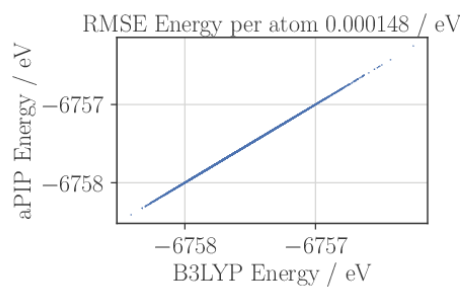

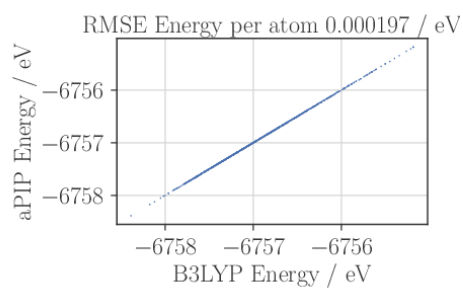

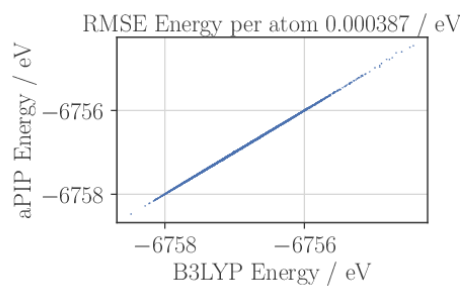

Figure S79: Comparison between the QM and PIP energies for NMA

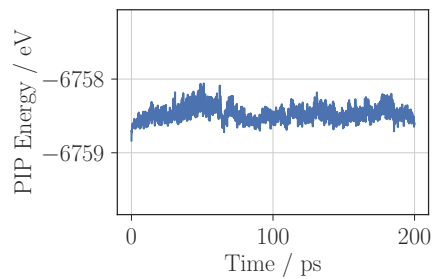

Figure S80: Energy over the course of a 300K MD simulation whilst using an aPIP potential for NMA

## S1.5   Energy per atom Errors

The energy per atom errors for the testing and training set forces are shown in Table S1.

Table S1: The energy per atom errors for the testing and training set for the molecules tested.

| Molecule | No. atoms | Energy RMSE per atom(meV) | | | | | |
|---|---|---|---|---|---|---|---|
| | | Training | | Testing 300K | | Testing High Temp. | |
| | | No Reg. | Reg. | No Reg. | Reg. | No Reg. | Reg. |
| Methane | 5 | 0.02 | 0.05 | 0.04 | 0.04 | 30.03 | 0.32 |
| Ethane | 8 | 0.06 | 0.23 | 0.06 | 0.15 | 3.26 | 0.98 |
| Propane | 11 | 0.21 | 0.28 | 0.20 | 0.27 | $5.64 \times 10^5$ | 1.15 |
| Butane | 14 | 0.35 | 0.58 | 0.47 | 0.50 | 19.45 | 2.08 |
| Pentane | 17 | 0.55 | 0.76 | 0.70 | 0.76 | 4.17 | 2.95 |
| Hexane | 20 | 0.78 | 0.98 | 1.26 | 1.41 | 3.69 | 3.29 |
| Adamantane | 26 | 0.33 | 0.44 | 0.13 | 0.16 | $2.83 \times 10^4$ | 0.85 |
| Ethene | 6 | 0.64 | 0.49 | 0.57 | 0.35 | $1.97 \times 10^3$ | 3.65 |
| Butene | 12 | 0.81 | 1.11 | 1.33 | 1.33 | 187.71 | 4.74 |
| Butadiene | 10 | 0.40 | 0.79 | 0.59 | 0.52 | $1.06 \times 10^6$ | 3.15 |
| Benzene | 12 | 0.16 | 0.36 | 0.15 | 0.16 | 8.07 | 0.74 |
| Methylbenzene | 15 | 0.32 | 0.57 | 0.22 | 0.27 | 9.05 | 1.71 |
| Ethanol | 9 | 0.61 | 0.36 | $1.50 \times 10^6$ | 0.16 | $3.13 \times 10^8$ | 0.62 |
| NMA | 12 | 0.21 | 0.33 | 0.21 | 0.15 | 4.11 | 0.39 |
| Mean | | 0.39 | 0.52 | $1.07 \times 10^5$ | 0.44 | $2.25 \times 10^7$ | 1.90 |

## S1.6 Force Errors

The force errors for the testing and training set forces are shown in Table S2.

Table S2: The force errors for the testing and training set for the molecules tested.

| Molecule | No. atoms | Force RMSE (meV/Å) | | | | | |
|---|---|---|---|---|---|---|---|
| | | Training | | Testing 300K | | Testing High Temp. | |
| | | No Reg. | Reg. | No Reg. | Reg. | No Reg. | Reg. |
| Methane | 5 | 0.59 | 2.86 | 1.01 | 1.2 | 2645 | 12.6 |
| Ethane | 8 | 3.07 | 23.7 | 1.89 | 13.5 | 224.6 | 57.7 |
| Propane | 11 | 15.0 | 25.4 | 11.1 | 12.3 | $1.89 \times 10^8$ | 49.3 |
| Butane | 14 | 34.4 | 49.2 | 33.1 | 27.8 | 2975 | 85.9 |
| Pentane | 17 | 49.6 | 61.4 | 38.4 | 39.0 | 198.8 | 112.3 |
| Hexane | 20 | 66.6 | 76.7 | 49.3 | 50.1 | 173.1 | 127.5 |
| Adamantane | 26 | 29.0 | 36.6 | 13.4 | 13.9 | $3.26 \times 10^6$ | 52.4 |
| Ethene | 6 | 11.4 | 46.6 | 13.1 | 30.5 | $8.00 \times 10^4$ | 10.8 |
| Butene | 12 | 60.9 | 109 | 66.9 | 53.7 | $2.39 \times 10^4$ | 161 |
| Butadiene | 10 | 32.8 | 61.0 | 41.1 | 29.3 | $9.37 \times 10^7$ | 110 |
| Benzene | 12 | 11.4 | 27.8 | 5.62 | 7.18 | 728.5 | 36.6 |
| Methylbenzene | 15 | 32.7 | 48.1 | 18.7 | 21.8 | 74.3 | 86.6 |
| Ethanol | 9 | 5.86 | 22.3 | $1.93 \times 10^{-8}$ | 10.3 | $1.93 \times 10^{10}$ | 34..1 |
| NMA | 12 | 12.0 | 20.3 | 13.1 | 9.86 | 455.9 | 22.8 |

## S1.7 MD Without Regularization

The MD energy trajectory for a butene aPIP potential with and without regularization is shown in Figure S81.

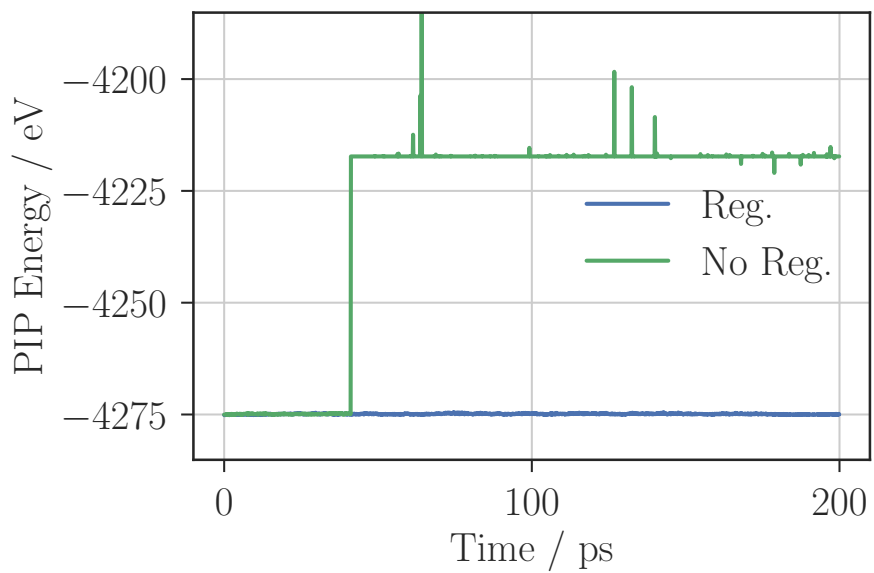

Figure S81: The energy of a butene molecule during a 300K MD simulation. An aPIP potential with regularization and a repulsive core is shown in comparison to an aPIP potential which is fit without regularization and a repulsive core.

# S2  Varying Polynomial Degree

In this section, the change in the performance of the potential with polynomial degree is shown for the molecule butene. Additionally, the performance of the potential if the bond length transform is used instead of the bond angle transform is also shown.

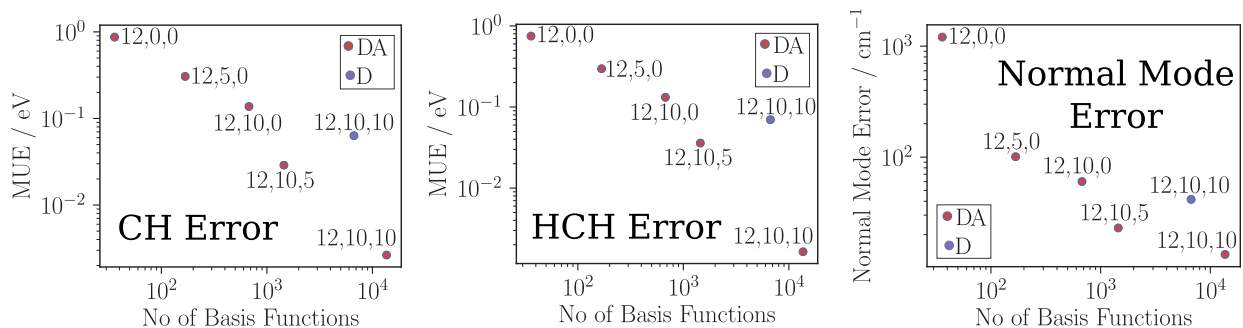

Figure S82: The change in the MUE for CH/HCH energy scans and the normal mode error with the degree employed in the polynomial. The degree of the polynomial is shown in the order 2B, 3B, 4B (12,0,0 therefore corresponds to a 2B potential, 12,10,0 corresponds to a 3B potential). The performance of the distance length (D) transform is also plotted.

The change in the performance of the potential with degree and transform is shown in Figure S82. As expected, with an increase in the body order and degree, the number of basis functions increase and the accuracy of the potential improves. The distance length (D) transform is seen to have a lower number of basis functions than the distance angle (DA) transform and a reduction in the performance. This is particularly noticeable for the recreation of the normal modes with the error for D, 41.6 cm$^{-1}$, three times higher than the DA transform with degree 12,10,10 (13.26 cm$^{-1}$) and almost double the DA potential with degree 12,10,5 (22.98 cm$^{-1}$). However, the decrease in performance is not simply due to the decrease in the number of basis functions as for all three properties the error for 12,10,5 (with 1448 basis functions) is below the D transform error (with 6681 basis functions).

Figure S83 and Figure S84 again show that the increase in the potential's degree leads to improved performance. A degree of 12,10,0 or 12,10,5 is seen to be insufficiently accurate to reproduce the dihedral scan and also results in a high error in the testing and training sets.

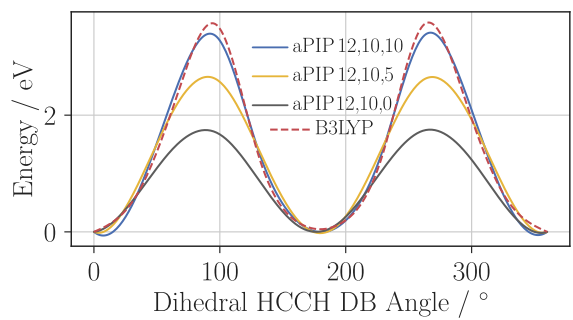

Figure S83: Energy curves for the H–C–C–H dihedral angle in butene. aPIPs with varying 4B degree are shown. The degree of the polynomial is shown in the order 2B, 3B, 4B.

Training Set - 1000 structures, 1500K MD

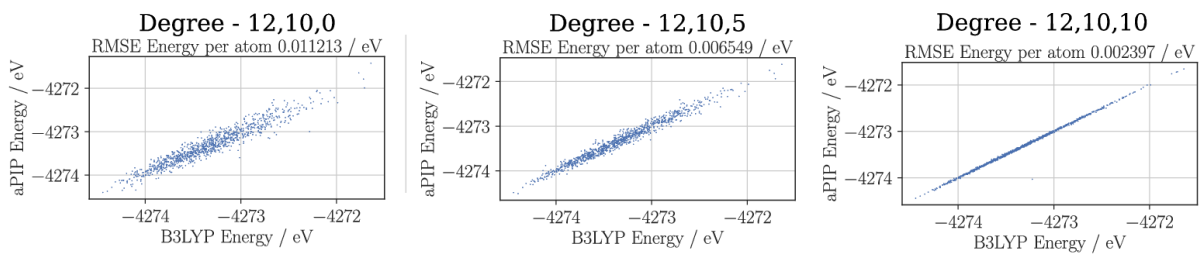

Testing Set - 8000 structures, 300K MD

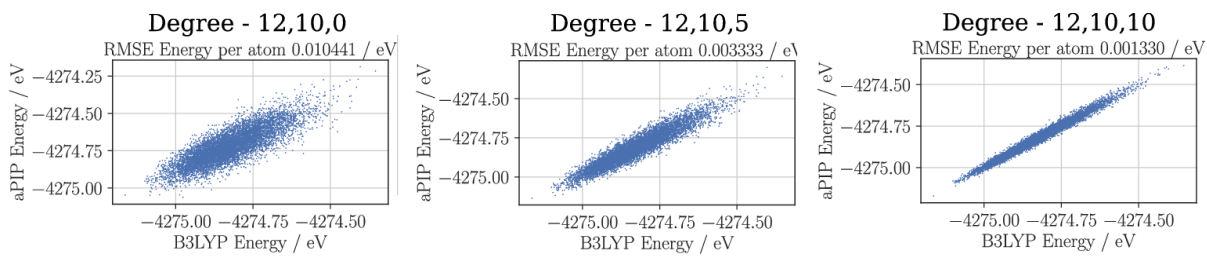

Testing Set - 8000 structures, 1500K MD

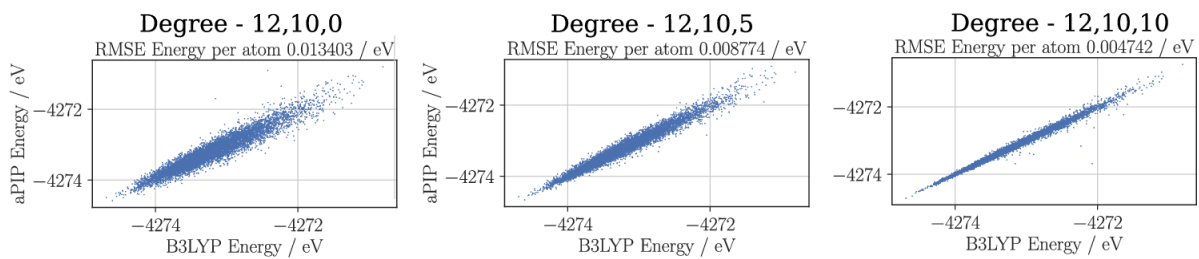

Figure S84: Comparison of the QM and PIP energies for Butene. aPIPs with varying 4B degree are shown. The degree of the polynomial is shown in the order 2B, 3B, 4B.

# S3 Testing the Speed of the Potential

The speed of three of the potentials created in this work was calculated by finding the energy and forces of 1000 structures. The table below summarizes the findings.

Table S3: The time taken per atom to calculate the energy and forces with aPIPs and with sGDML. The sGDML - Intel Xeon CPU E5-2640 @ 2.40 GHz timings are taken from Ref. S1.

| | Timings (ms/per atom) | | | |
|---|---|---|---|---|
| | **Run 1** | **Run 2** | **Run 3** | **Mean** |
| *aPIP - Intel Xeon CPU E5-2680 @ 2.40GHz* | | | | |
| **Methane** | 0.106 | 0.0853 | 0.0919 | 0.0944 |
| **Butane** | 0.741 | 0.748 | 0.728 | 0.739 |
| **Benzene** | 0.647 | 0.673 | 0.63 | 0.65 |
| **Ethanol** | 0.795 | 0.822 | 0.828 | 0.815 |
| *sgdml - Intel Xeon CPU E5-2640 @ 2.40 GHz* [a] | | | | |
| **Benzene** | | | | 0.192 |
| **Ethanol** | | | | 0.134 |
| *sgdml - Intel Xeon CPU E5-2680 @ 2.40GHz* | | | | |
| **Benzene** | 0.367 | 0.365 | 0.405 | 0.379 |
| **Ethanol** | 0.188 | 0.193 | 0.186 | 0.189 |

Table S3 shows that all three molecules have a speed of less than 1.0 ms/per atom. The three hydrocarbon molecules demonstrate the increase in time with the number of atoms present in the molecule, this is because there is an increase in time required with the number of atoms in the cutoff. This trend of speed with size does not persist to ethanol, which is slower than benzene despite consisting of 9 atoms as opposed to 12 atoms. This is due to ethanol consisting of three different elements and therefore having a greater number of basis functions (27437) than the hydrocarbon potentials (13629 for benzene).

The aPIPs benzene and ethanol potentials are several times slower than the sGDML potentials. However, given that there are further opportunities to improve the speed of aPIPs in future versions of the potential, this discrepancy in the timings is not seen as a concern.

# S4 Distance-based coordinates invariants

We present below primary and secondary invariants for three-body and four-body potentials with distance-based coordinates.

Table S4: 3-body primary invariants for distance-based coordinates. For this body-order there is only the trivial secondary invariant $s_1 = 1$.

|       | AAA                                      | ABB             | ABC      |
|-------|------------------------------------------|-----------------|----------|
| $p_1$ | $u_{12} + u_{13} + u_{23}$               | $u_{12} + u_{13}$ | $u_{12}$ |
| $p_2$ | $u_{12}u_{13} + u_{12}u_{23} + u_{13}u_{23}$ | $u_{12}u_{13}$  | $u_{13}$ |
| $p_3$ | $u_{12}u_{13}u_{23}$                     | $u_{23}$        | $u_{23}$ |
| $s_1$ | 1                                        | 1               | 1        |

Table S5: 4-body primary and secondary invariants for distance-based coordinates.

|       | ABBB                                                                                  | AABB                                       | AABC                              | ABCD     |
|-------|---------------------------------------------------------------------------------------|--------------------------------------------|-----------------------------------|----------|
| $p_1$ | $u_{23} + u_{24} + u_{34}$                                                             | $u_{12}$                                   | $u_{12}$                          | $u_{12}$ |
| $p_2$ | $u_{34}$                                                                               | $u_{34}$                                   | $u_{13}$                          |          |
| $p_3$ | $u_{23}^3 + u_{24}^3 + u_{34}^3$                                                       | $u_{13} + u_{14} + u_{23} + u_{24}$        | $u_{13} + u_{23}$                 | $u_{14}$ |
| $p_4$ | $u_{12} + u_{13} + u_{14}$                                                             | $u_{13}u_{14} + u_{23}u_{24}$              | $u_{14} + u_{24}$                 | $u_{23}$ |
| $p_5$ | $u_{12}^2 + u_{13}^2 + u_{14}^2$                                                       | $u_{13}u_{23} + u_{14}u_{4}$               | $u_{13}^2 + u_{23}^2$             | $u_{24}$ |
| $p_6$ | $u_{12}^3 + u_{13}^3 + u_{14}^3$                                                       | $u_{13}^2 + u_{14}^2 + u_{23}^2 u_{24}^2$  | $u_{14}^2 + u_{24}^2$             | $u_{34}$ |
| $s_1$ | 1                                                                                     | 1                                          | 1                                 | 1        |
| $s_2$ | $u_{23}u_{14} + u_{24}u_{13} + u_{34}u_{12}$                                           | $u_{13}^3 + u_{14}^3 + u_{23}^3 + u_{24}^3$ | $u_{13}u_{14} + u_{23}u_{24}$    |          |
| $s_3$ | $s_2^2$                                                                                |                                            |                                   |          |
| $s_4$ | $s_2^3$                                                                                |                                            |                                   |          |
| $s_5$ | $u_{23}u_{24}(u_{13} + u_{14}) + u_{23}u_{34}(u_{12} + u_{14}) + u_{24}u_{34}(u_{12} + u_{13})$ | |                                   |          |
| $s_6$ | $u_{23}u_{14}(u_{12} + u_{13}) + u_{24}u_{13}(u_{12} + u_{14}) + u_{34}u_{12}(u_{13} + u_{14})$ | |                                   |          |




%
%
%
%

\end{document}